\documentclass[a4paper]{article}
\usepackage{amsmath}
\usepackage{amsfonts}
\usepackage{amsthm}
\usepackage{amsopn}
\usepackage{graphicx}
\usepackage{tikz}
\usepackage{float}

\def\be{\begin{eqnarray}}
\def\ee{\end{eqnarray}}
\def\nn{\nonumber}

\def\p{\partial}
\def\tr{{\rm tr}\,}
\def\Tr{{\rm tr}\,}

\def\l[{\phantom.[}

\hoffset= - 1.0in         

\begin{document}

\title{{\bf {On matrix-model approach to simplified Khovanov-Rozansky calculus
}\vspace{.2cm}}
\author{{\bf A.Morozov$^{a,b,c}$},\ {\bf And.Morozov$^{a,b,c,d}$} \ and \ {\bf A.Popolitov$^{a,b,e}$}}
\date{ }
}

\maketitle

\vspace{-5.5cm}

\begin{center}
\hfill IITP/TH-08/15\\
\hfill ITEP/TH-15/15\\
\end{center}

\vspace{3.3cm}

\begin{center}

$^a$ {\small {\it ITEP, Moscow 117218, Russia}}\\
$^b$ {\small {\it Institute for Information Transmission Problems, Moscow 127994, Russia}}\\
$^c$ {\small {\it National Research Nuclear University MEPhI, Moscow 115409, Russia }}\\
$^d$ {\small {\it Laboratory of Quantum Topology, Chelyabinsk State University, Chelyabinsk 454001, Russia}} \\
$^e$ {\small {\it Korteweg-de Vries Institute for Mathematics, University of
Amsterdam, P.O. Box 94248, 1090 GE Amsterdam, The Netherlands}}
\end{center}


\vspace{1cm}

\centerline{ABSTRACT}

\bigskip

{\footnotesize
Wilson-loop averages in Chern-Simons theory (HOMFLY polynomials) can be evaluated
in different ways -- the most difficult, but most interesting of them is the
hypercube calculus, the only one applicable to virtual knots
and used also for categorification (higher-dimensional extension) of the theory.
We continue the study of quantum dimensions, associated with hypercube vertices,
in the drastically simplified version of this approach to knot polynomials.
At $q=1$ the problem is reformulated in terms of fat (ribbon) graphs,
where Seifert cycles play the role of vertices.
Ward identities in associated matrix model
provide a set of recursions between classical dimensions.
For $q\neq 1$ most of these relations are broken
(i.e. deformed in a still uncontrollable way),
and only few are protected by Reidemeister invariance of Chern-Simons theory.
Still they are helpful for systematic evaluation of entire series of quantum dimensions,
including negative ones, which are relevant for virtual link diagrams.
To illustrate the effectiveness of developed formalism we derive explicit expressions
for the 2-cabled HOMFLY of virtual trefoil and virtual 3.2 knot, which involve respectively
12 and 14 intersections -- far beyond any dreams with alternative methods.
As a more conceptual application, we describe a
relation between the genus of fat graph and Turaev genus of original
link diagram, which is currently the most effective tool for the search of
thin knots.
}

\vspace{1cm}

$3d$ Chern-Simons theory (CST) \cite{CS} is the next exactly-solvable model which should be
exhaustively understood after the recent breakthrough in $2d$ conformal theory \cite{CFTAGT}.
The task of CST is to provide clear and technically effective description
of observables, which are Wilson-loop averages along various contours
in various representations, also known as colored knot/link polynomials \cite{knotpols}.
Among numerous remaining puzzles in CST  the two biggest are the $\beta$-deformation \cite{betadefo}
(the theory of Khovanov-Rozansky \cite{KhR}, super-polynomials \cite{superpols}
and Kapustin-Witten equations \cite{KW})
and the extension of the most powerful Reshetikhin-Turaev (RT) approach \cite{RT}-\cite{AnoMcabling}
from ordinary to virtual \cite{virt} knots and links.

Remarkably, it looks like both problems can be attacked simultaneously --
at least a very promising approach to both is provided by the recent suggestion
in \cite{DM3} -- see  \cite{AnoMKhR} and \cite{DM3virt}.
Suggestion is to directly generalize the hypercube construction of M.Khovanov
and D.Bar-Natan \cite{Kh}
from $sl(2)$, when it could be related to the ${\cal R}$-matrix of
L.Kauffman \cite{KauffR}, to $sl(N)$, when such ${\cal R}$-matrix is unknown
-- and do this without overcomplicated matrix-factorization technique, used in \cite{KhR}.
The price to pay is a little more sophisticated pattern of spaces, associated
with hypercube vertices, which can now be factor-spaces and even (in the case
of virtual knots) have negative dimensions, i.e. need to be treated by $K$-theory methods.
The purpose of this paper is to move further in determining these dimensions
and extend the list, known from \cite{DM3,AnoMKhR} and \cite{DM3virt}.
It turns out that natural language for the problem is the one of matrix models and fat graphs,
as was partly anticipated in \cite{fatgraphs}.

\section{Simplified substitute of Khovanov-Rozansky formalism}

In the present text we assume familiarity with the hypercube construction of \cite{Kh}
and just describe the peculiarities coming from the suggestion of \cite{DM3}.

The starting point  is a link diagram ${\cal L}$, which is a graph with $n$ four-valent
vertices of two colors, with which one associates a hypercube ${\cal H}^{\cal L}$,
where each of $2^n$ vertices is associated with particular choice of coloring ${\cal L}_c$.
The {\it base} (or {\it Seifert}) vertex of ${\cal H}^{\cal L}$ is the one where all vertex colors are black,
and link diagram is resolved into collection of {\it Seifert cycles}.

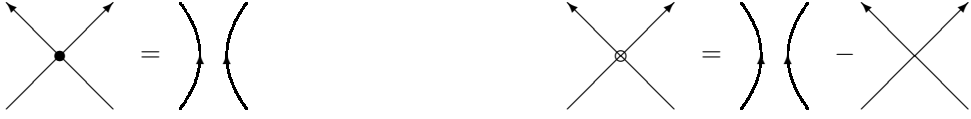
\begin{picture}(300,60)(-130,-30)
\put(-80,-20){\vector(1,1){40}}
\put(-40,-20){\vector(-1,1){40}}
\put(-60,0){\circle*{4}}
\put(-30,-2){\mbox{$=$}}
\linethickness{0.2mm}
\qbezier(-15,-20)(0,0)(-15,20)
\qbezier(10,-20)(-5,0)(10,20)
\put(-7.5,-1){\vector(0,1){2}}
\put(2.5,-1){\vector(0,1){2}}
\put(130,-20){\vector(1,1){40}}
\put(170,-20){\vector(-1,1){40}}
\put(150,0){\circle{4}}
\put(180,-2){\mbox{$=$}}
\qbezier(195,-20)(210,0)(195,20)
\qbezier(220,-20)(205,0)(220,20)
\put(202.5,-1){\vector(0,1){2}}
\put(212.5,-1){\vector(0,1){2}}
\put(230,-2){\mbox{$-$}}
\put(240,-20){\vector(1,1){40}}
\put(280,-20){\vector(-1,1){40}}
\end{picture}

\noindent
The peculiarity of \cite{DM3} is that white resolution is associated with
the difference $\Pi$ between identity and simple crossing
(instead of the "horizontal" resolution in \cite{KauffR,Kh}, which does not respect
arrows and therefore is applicable only at $N=2$).
It is this {\it difference} that is responsible for factor-spaces and potential
negativity.

At particular vertex of $v\in {\cal H}^{\cal L}$ some vertices of ${\cal L}$ are black
and some are white, and this converts ${\cal L}$ into an algebraic sum of closed cycles.
Classically (at $q=1$) one substitutes each cycle by $N$ and therefore associated
with $v$ is a polynomial in $N$.
At $q\neq 1$ it is not enough to substitute $N$ by quantum $[N]=\frac{q^N-q^{-N}}{q-q^{-1}}$
(what was sufficient in the case of $N=2$ in \cite{KauffR,Kh}) --
this polynomial is rather a $q$-deformed (quantum) dimension (perhaps, negative),
which was technically defined in \cite{AnoMKhR} and
which we are going to describe better in the present paper.
As explained in \cite{DM3}, quantization procedure is very much restricted,
because different colorings of ${\cal L}$ provide different knots, while the hypercube
and dimensions are the same.
This allows to proceed recursively -- from simpler to more complicated knots
and reconstruct dimensions from topological invariance.
Our goal here is to look for a better and more algorithmic formulation
of such recursion, and understand possible subtleties and loopholes.

The $q$-alternated sum of these dimensions,
\be
H_{_\Box}^{{\cal L}_c} \ = \  \sum_{v\in {\cal H}^{\cal L}} \ (-q)^{h_v-h_c}\cdot d_v(N,q)
\label{qaltsum}
\ee
defines HOMFLY polynomial for ${\cal L}$ with a given coloring $c$
(which enters the sum through the choice of "initial vertex" $v_c$ and the "height"
$h_v$ is the minimal number of hypercube edges in between $v$ and the Seifert vertex $v_S$).
Khovanov-Rozansky polynomial \cite{KhR} is made in a similar way from cohomologies
of abelian complex, to which hypercube is converted when certain commuting
morphisms are defined along the hypercube's edges.
Understanding "dimensions" of graded factor-spaces at hypercube vertices
is the necessary step in construction
of these morphisms -- and it is the step that we concentrate on in the present paper.

We study the interplay and interrelation between four different approaches to
evaluation of quantum dimensions:

$\bullet$ recursive definition from the change of coloring \cite{DM3}

$\bullet$ direct evaluation from a combination of projectors $\pi_{_{[11]}}$ \cite{AnoMKhR}

$\bullet$ matrix-model-inspired recursion [this paper]

$\bullet$ phenomenological rules in the spirit of evolution method of \cite{evo}

\noindent
In result we obtain a rather effective and computerizable calculational machinery,
which is applied to the study of topological invariance and to evaluation of the first
cabled HOMFLY for virtual knots.

\bigskip

{\bf The simplest examples and notation:}

\bigskip

{\bf Ex.1.} Our first example is the  {eight-diagram ${\cal L}= \bigcirc\!\!\!\bullet\!\!\!\bigcirc$},
which is one of realizations of the unknot.
It has one vertex, thus there are two different resolutions of ${\cal L}$
and the corresponding hypercube ${\cal H}^{\cal L}$ has two -- it is just a segment.
The base (Seifert) resolution provides two Seifert cycles, another resolution is a difference.
If we denote each Seifert cycle by a solid circle, then we get two disconnected vertices
over the Seifert vertex $v_S\in{\cal H}^{\cal L}$ and two, connected by a propagator $\Pi$ over
another hypercube vertex.

The propagator
\be
<M^i_j\ M^k_l>\ =\ \Pi^{ik}_{jl} = \delta^i_j\delta^k_l - \delta^i_l\delta^k_j,
\ \ \ \ \ \ \ \ i,j,k,l = 1,\ldots, N
\label{claprop}
\ee
denotes the twisted (Mobius-like) strip

\begin{picture}(300,55)(-100,-17)
  \thicklines
  \put(0,0){\vector(0,1){20}}
      {\linethickness{0.15mm} \qbezier(0,20)(40,20)(50,0) \qbezier(0,0)(40,0)(50,20) }
  \put(50,0){\vector(0,1){20}}
  \put(11,7){\mbox{$\Pi$}} \put(-14,7){\mbox{M}} \put(54,7){\mbox{M}} \put(70,7){\mbox{$=$}}
 \put(5,0){
  \put(-2,22){\mbox{i}} \put(-2,-10){\mbox{j}} \put(48,22){\mbox{k}} \put(48,-10){\mbox{l}}
  \put(90,0) {
    \put(0,0){\vector(0,1){20}}
    \put(30,0){\vector(0,1){20}}
    \put(-2,22){\mbox{i}} \put(-2,-10){\mbox{j}} \put(28,22){\mbox{k}} \put(28,-10){\mbox{l}}
    \put(-9,7){$\delta$} \put(34,7){$\delta$}
  }
  \put(140,7){\mbox{$-$}}
  \put(170,0) {
    \put(50,0){\vector(-3,1){50}} \put(0,0){\vector(3,1){50}}
    \put(-2,22){\mbox{i}} \put(-2,-10){\mbox{j}} \put(48,22){\mbox{k}} \put(48,-10){\mbox{l}}
    \put(21,21){$\delta$}     \put(21,-10){$\delta$}
  }}
\end{picture}

\noindent
which in this particular case can be untwisted by changing mutual orientation (vorticities)
of the two vertices.

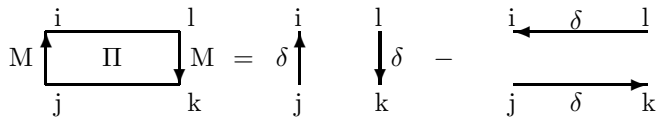
\begin{picture}(300,45)(-100,-10)
  \thicklines
  \put(0,0){\vector(0,1){20}}
  {\thinlines \put(0,20){\line(1,0){50}} \put(0,0){\line(1,0){50}}}
  \put(50,20){\vector(0,-1){20}}
  \put(21,7){\mbox{$\Pi$}} \put(-14,7){\mbox{M}} \put(54,7){\mbox{M}} \put(70,7){\mbox{$=$}}
\put(5,0){
  \put(-2,22){\mbox{i}} \put(-2,-10){\mbox{j}} \put(48,22){\mbox{l}} \put(48,-10){\mbox{k}}
  \put(90,0) {
    \put(0,0){\vector(0,1){20}}
    \put(30,20){\vector(0,-1){20}}
    \put(-2,22){\mbox{i}} \put(-2,-10){\mbox{j}} \put(28,22){\mbox{l}} \put(28,-10){\mbox{k}}
    \put(-9,7){$\delta$} \put(34,7){$\delta$}
  }
  \put(140,7){\mbox{$-$}}
  \put(170,0) {
    \put(50,20){\vector(-1,0){50}} \put(0,0){\vector(1,0){50}}
    \put(-2,22){\mbox{i}} \put(-2,-10){\mbox{j}} \put(48,22){\mbox{l}} \put(48,-10){\mbox{k}}
    \put(21,21){$\delta$}     \put(21,-10){$\delta$}
  }}
\end{picture}

\bigskip

\newpage
Coming back to our eight-shaped unknot, we have the following hypercube (segment) of (two) resolutions:

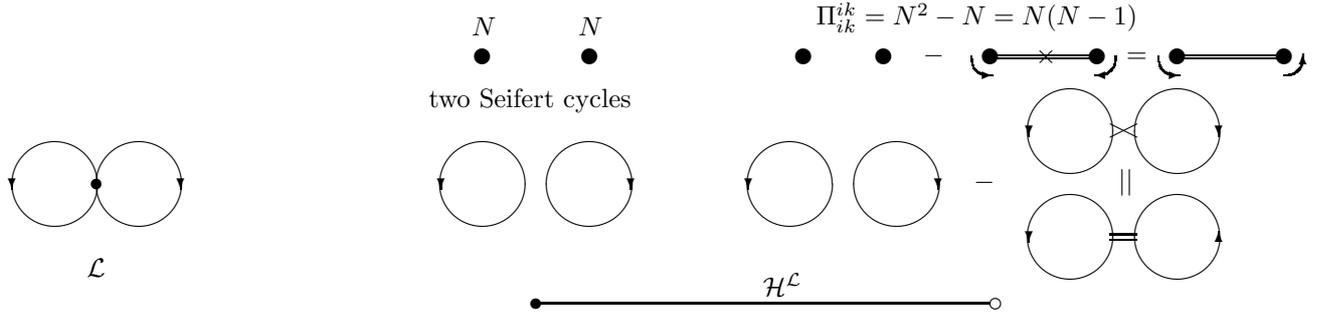
\begin{picture}(300,150)(-15,-70)
\put(0,0){\circle{30}}\put(31.5,0){\circle{30}}\put(15.5,0){\circle*{4}}
\put(-16,0){\vector(0,-1){2}}\put(47,0){\vector(0,-1){2}}
\put(12,-35){\mbox{${\cal L}$}}
\put(160,0){\circle{30}} \put(200,0){\circle{30}}
\put(144.5,0){\vector(0,-1){2}}\put(215.5,0){\vector(0,-1){2}}
\put(180,-45){\circle*{4}}
\put(-15,0){
\put(290,0){\circle{30}} \put(330,0){\circle{30}}
\put(274.5,0){\vector(0,-1){2}}\put(345.5,0){\vector(0,-1){2}}
}
\put(344,-2){\mbox{$-$}}
%
\put(380,20){\circle{30}} \put(420,20){\circle{30}}
\put(364.5,20){\vector(0,-1){2}}\put(435.5,20){\vector(0,-1){2}}
\put(395,17.5){\line(2,1){10}}\put(395,22.5){\line(2,-1){10}}
\put(398,-2){\mbox{$||$}}
\put(380,-20){\circle{30}} \put(420,-20){\circle{30}}
\put(364.5,-20){\vector(0,-1){2}}\put(435.5,-20){\vector(0,1){2}}
\put(395,-21){\line(1,0){10}}\put(395,-19){\line(1,0){10}}
\put(352,-45){\circle{4}}
\put(180,-45){\line(1,0){170}}
\put(265,-42){\mbox{${\cal H}^{\cal L}$}}
\put(160,48){\circle*{6}}\put(200,48){\circle*{6}}
\put(280,48){\circle*{6}}\put(310,48){\circle*{6}}
\put(325,46){\mbox{$-$}}
\put(60,0){
\put(290,48){\circle*{6}}\put(330,48){\circle*{6}}
\qbezier(283,48)(283,41)(290,41)\put(289,41){\vector(1,0){2}}
\qbezier(337,48)(337,41)(330,41)\put(331,41){\vector(-1,0){2}}
\put(290,47.5){\line(1,0){40}}\put(290,48.5){\line(1,0){40}}
\put(307,45.5){\mbox{$\times$}}
\put(341,46){\mbox{$=$}}
\put(360,48){\circle*{6}}\put(400,48){\circle*{6}}
\qbezier(353,48)(353,41)(360,41)\put(359,41){\vector(1,0){2}}
\qbezier(407,48)(407,41)(400,41)\put(407,47){\vector(0,1){2}}
\put(360,47.5){\line(1,0){40}}\put(360,48.5){\line(1,0){40}}
}
\put(140,29){\mbox{two Seifert cycles}}
\put(156,56){\mbox{$N$}}\put(196,56){\mbox{$N$}}
\put(285,60){\mbox{$\Pi^{ik}_{ik} =N^2-N =N(N-1)$}}
\end{picture}

\noindent
The two dimensions at hypercube vertices are, classically,
$d_S = d_\bullet = N^2$ and $d_\circ = N(N-1)$, and at $q\neq 1$ they are
deformed (quantized) to become
\be
d_\bullet^{\infty} = [N]^2, \ \ \ \ \ \ \ \ \  d_\circ^{\infty} = [N][N-1]
\ee
where $[x] = \frac{\{q^x\}}{\{q\}} = \frac{q^x-q^{-x}}{q-q^{-1}}$ is the quantum number
and $\{x\} = x-x^{-1}$.
According to the general rules of Khovanov calculus \cite{DM3}
the unreduced HOMFLY polynomial in the fundamental representation is either
\be
H^{\infty}_{_\Box} \stackrel{(\ref{qaltsum})}{\equiv}
q^{N-1}\Big(d_\bullet^{\infty}-q\cdot d_\circ^{\infty}\Big) =
q^{N-1}\Big([N]^2 - q[N][N-1]\Big) = [N]
\ee
or
\be
H^{\infty}_{_\Box} \stackrel{(\ref{qaltsum})}{\equiv}
-q^{-N}\Big(d_\circ^{\infty}-q\cdot d_\bullet^{\infty}\Big) =
-q^{-N}\Big([N][N-1]-q[N]^2\Big) = [N]
\ee
as requested for the unknot.

\bigskip

{\bf Ex.2.} There is just one another example of link diagram with a single vertex and thus
with a one-dimensional hypercube: it is the \underline{virtual Hopf link}

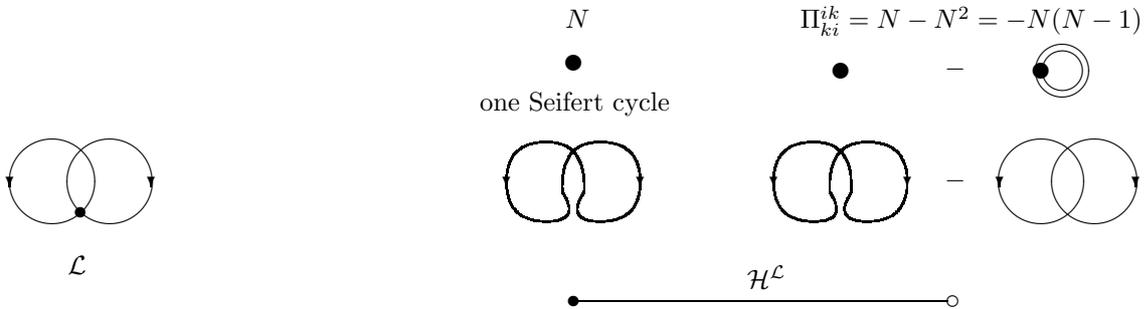
\begin{picture}(300,140)(-110,-65)
  \linethickness{0.2mm}
\put(-90,0){\circle{30}}\put(-68.5,0){\circle{30}}\put(-79.5,-11.5){\circle*{4}}
\put(-106,0){\vector(0,-1){2}}\put(-53,0){\vector(0,-1){2}}
\put(-84,-35){\mbox{${\cal L}$}}
\qbezier(80,0)(80,15)(95,15)\qbezier(80,0)(80,-15)(95,-15)
\qbezier(109,-4)(110,15)(95,15)\qbezier(95,-15)(108,-15)(101,-4)
\qbezier(101,-4)(100,15)(115,15)\qbezier(115,-15)(102,-15)(109,-4)
\qbezier(130,0)(130,15)(115,15)\qbezier(130,0)(130,-15)(115,-15)
\put(80,0){\vector(0,-1){2}}\put(130,0){\vector(0,-1){2}}
\qbezier(180,0)(180,15)(195,15)\qbezier(180,0)(180,-15)(195,-15)
\qbezier(209,-4)(210,15)(195,15)\qbezier(195,-15)(208,-15)(201,-4)
\qbezier(201,-4)(200,15)(215,15)\qbezier(215,-15)(202,-15)(209,-4)
\qbezier(230,0)(230,15)(215,15)\qbezier(230,0)(230,-15)(215,-15)
\put(180,0){\vector(0,-1){2}}\put(230,0){\vector(0,-1){2}}
\put(244,-2){\mbox{$-$}}
\put(280,0){\circle{30}} \put(300,0){\circle{30}}
\put(264.5,0){\vector(0,-1){2}}\put(315.5,0){\vector(0,-1){2}}
%
%
\put(105,-45){\circle*{4}} \put(247,-45){\circle{4}}
\put(105,-45){\line(1,0){140}}
\put(170,-40){\mbox{${\cal H}^{\cal L}$}}
\put(105,45){\circle*{6}}
\put(70,27){\mbox{one Seifert cycle}}
\put(102,58){\mbox{$N$}}
\put(205,42){\circle*{6}}
\put(244,40){\mbox{$-$}}
\put(280,42){\circle*{6}}
\put(288,42){\circle{15}}
\put(288,42){\circle{18}}
\put(190,58){\mbox{$\Pi^{ik}_{ki} = N-N^2 = -N(N-1)$}}
\end{picture}

\noindent
In this case there is one real intersection, denoted by dot, while another one
is "sterile" -- appears as an artefact of projection from non-simply-connected
space-time to two-dimensional plane.
Sometimes we put sterile crossings into big circles, to emphasize what they are.

Sterile intersections affect the shape and the number of cycles.
That is why, in the case of the virtual Hopf
there is just one Seifert cycle, and at the second vertex of the hypercube
the would-be "dimension" is negative(!)
[{\bf NB:} Note that twisting of the edge is now crucially important -- if untwisted, the
edge would contribute $-N^2$ instead of $-N$.]
After quantization we get
\be
d_S^{\rm vHopf} = d_\bullet^{\rm vHopf} = [N], \ \ \ \ \ \ \ \
d_\circ^{\rm vHopf} = -[N][N-1]
\ee
and the unreduced fundamental HOMFLY polynomial is either
\be
H^{\rm vHopf}_{_\Box}(q) \stackrel{\cite{DM3}}{\equiv}
q^{N-1}\Big(d_\bullet^{\rm vHopf}-q\cdot d_\circ^{\rm vHopf}\Big) =
q^{N-1}\Big([N] + q[N][N-1]\Big) = [N]\Big(q^N[N-1]+q^{N-1}\Big)
\ee
or
\be
H^{\rm vHopf}_{_\Box}(q^{-1}) \stackrel{\cite{DM3}}{\equiv}
-q^{-N}\Big(d_\circ^{\rm vHopf}-q\cdot d_\bullet^{\rm vHopf}\Big) =
q^{-N}\Big([N][N-1]+q[N]\Big) = [N]\Big(q^{-N}[N-1]+q^{1-N}\Big)
\ee

\bigskip

{\bf Ex.3.} For the ordinary \underline{Hopf link}

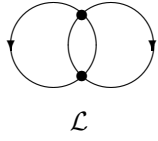
\begin{picture}(300,65)(-200,-40)
\put(-90,0){\circle{30}}\put(-68.5,0){\circle{30}}
\put(-79.5,-11.5){\circle*{4}}\put(-79.5,11.5){\circle*{4}}
\put(-106,0){\vector(0,-1){2}}\put(-53,0){\vector(0,-1){2}}
\put(-84,-32){\mbox{${\cal L}$}}
\end{picture}

\noindent
both intersections are real,
the hypercube is two-dimensional (a square with four vertices),
there are two Seifert cycles and all dimensions are positive:

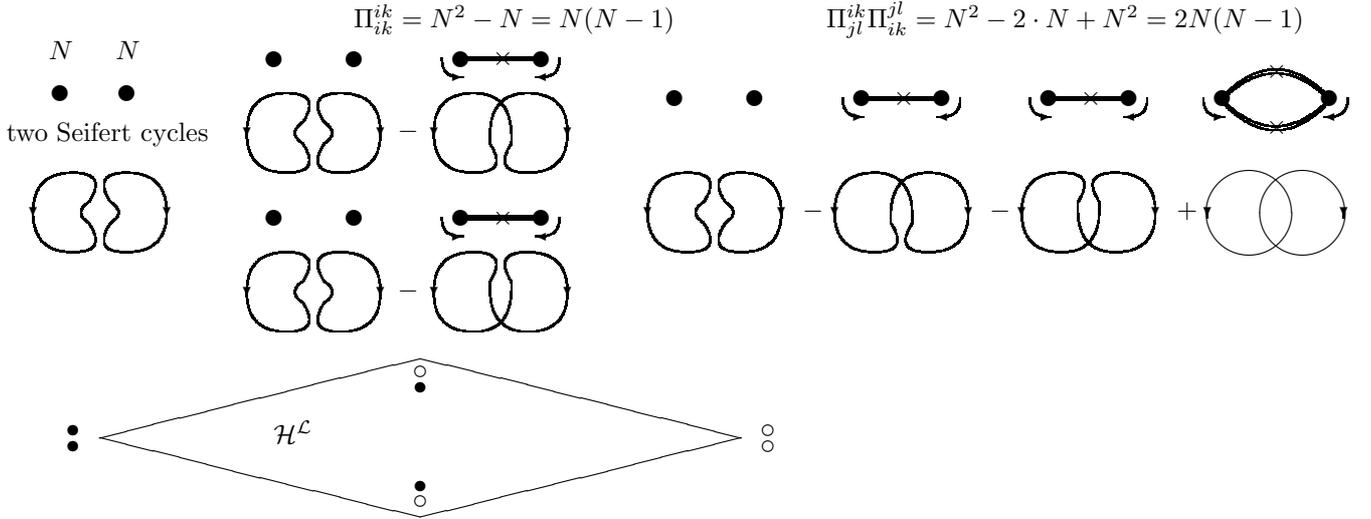
\begin{picture}(300,220)(90,-125)
    \linethickness{0.2mm}
  %
%
%
%
\qbezier(80,0)(80,15)(95,15)\qbezier(80,0)(80,-15)(95,-15)
\qbezier(101,4)(108,15)(95,15)\qbezier(95,-15)(108,-15)(101,-4)
\qbezier(101,4)(95,0)(101,-4)\qbezier(109,4)(115,0)(109,-4)
\qbezier(109,4)(102,15)(115,15)\qbezier(115,-15)(102,-15)(109,-4)
\qbezier(130,0)(130,15)(115,15)\qbezier(130,0)(130,-15)(115,-15)
\put(80,0){\vector(0,-1){2}}\put(130,0){\vector(0,-1){2}}
\qbezier(160,30)(160,45)(175,45)\qbezier(160,30)(160,15)(175,15)
\qbezier(181,34)(188,45)(175,45)\qbezier(175,15)(188,15)(181,26)
\qbezier(181,34)(175,30)(181,26)\qbezier(189,34)(195,30)(189,26)
\qbezier(189,34)(182,45)(195,45)\qbezier(195,15)(182,15)(189,26)
\qbezier(210,30)(210,45)(195,45)\qbezier(210,30)(210,15)(195,15)
\put(160,30){\vector(0,-1){2}}\put(210,30){\vector(0,-1){2}}
\qbezier(230,30)(230,45)(245,45)\qbezier(230,30)(230,15)(245,15)
\qbezier(259,26)(260,45)(245,45)\qbezier(245,15)(258,15)(251,26)
\qbezier(251,26)(250,45)(265,45)\qbezier(265,15)(252,15)(259,26)
\qbezier(280,30)(280,45)(265,45)\qbezier(280,30)(280,15)(265,15)
\put(230,30){\vector(0,-1){2}}\put(280,30){\vector(0,-1){2}}
\put(216.5,28){\mbox{$-$}}
\qbezier(160,-30)(160,-45)(175,-45)\qbezier(160,-30)(160,-15)(175,-15)
\qbezier(181,-34)(188,-45)(175,-45)\qbezier(175,-15)(188,-15)(181,-26)
\qbezier(181,-34)(175,-30)(181,-26)\qbezier(189,-34)(195,-30)(189,-26)
\qbezier(189,-34)(182,-45)(195,-45)\qbezier(195,-15)(182,-15)(189,-26)
\qbezier(210,-30)(210,-45)(195,-45)\qbezier(210,-30)(210,-15)(195,-15)
\put(160,-30){\vector(0,-1){2}}\put(210,-30){\vector(0,-1){2}}
\qbezier(230,-30)(230,-45)(245,-45)\qbezier(230,-30)(230,-15)(245,-15)
\qbezier(259,-26)(260,-45)(245,-45)\qbezier(245,-15)(258,-15)(251,-26)
\qbezier(251,-26)(250,-45)(265,-45)\qbezier(265,-15)(252,-15)(259,-26)
\qbezier(280,-30)(280,-45)(265,-45)\qbezier(280,-30)(280,-15)(265,-15)
\put(230,-30){\vector(0,-1){2}}\put(280,-30){\vector(0,-1){2}}
\put(216.5,-32){\mbox{$-$}}
\qbezier(310,0)(310,15)(325,15)\qbezier(310,0)(310,-15)(325,-15)
\qbezier(331,4)(338,15)(325,15)\qbezier(325,-15)(338,-15)(331,-4)
\qbezier(331,4)(325,0)(331,-4)\qbezier(339,4)(345,0)(339,-4)
\qbezier(339,4)(332,15)(345,15)\qbezier(345,-15)(332,-15)(339,-4)
\qbezier(360,0)(360,15)(345,15)\qbezier(360,0)(360,-15)(345,-15)
\put(310,0){\vector(0,-1){2}}\put(360,0){\vector(0,-1){2}}
\put(368,-2){\mbox{$-$}}
\qbezier(380,0)(380,15)(395,15)\qbezier(380,0)(380,-15)(395,-15)
\qbezier(409,-4)(410,15)(395,15)\qbezier(395,-15)(408,-15)(401,-4)
\qbezier(401,-4)(400,15)(415,15)\qbezier(415,-15)(402,-15)(409,-4)
\qbezier(430,0)(430,15)(415,15)\qbezier(430,0)(430,-15)(415,-15)
\put(380,0){\vector(0,-1){2}}\put(430,0){\vector(0,-1){2}}
\put(438,-2){\mbox{$-$}}
\qbezier(450,0)(450,15)(465,15)\qbezier(450,0)(450,-15)(465,-15)
\qbezier(479,4)(480,-15)(465,-15)\qbezier(465,15)(478,15)(471,4)
\qbezier(471,4)(470,-15)(485,-15)\qbezier(485,15)(472,15)(479,4)
\qbezier(500,0)(500,15)(485,15)\qbezier(500,0)(500,-15)(485,-15)
\put(450,0){\vector(0,-1){2}}\put(500,0){\vector(0,-1){2}}
\put(508,-2){\mbox{$+$}}
\put(535,0){\circle{30}} \put(555,0){\circle{30}}
\put(519.5,0){\vector(0,-1){2}}\put(570.5,0){\vector(0,-1){2}}
\put(90,45){\circle*{6}} \put(115,45){\circle*{6}}
\put(70,27){\mbox{two Seifert cycles}}
\put(86,58){\mbox{$N$}} \put(111,58){\mbox{$N$}}
\put(35,0){
\put(205,58){\circle*{6}}\put(235,58){\circle*{6}}
\qbezier(198,58)(198,51)(205,51)\put(204,51){\vector(1,0){2}}
\qbezier(242,58)(242,51)(235,51)\put(235,51){\vector(-1,0){2}}
\put(205,57.5){\line(1,0){30}}\put(205,58.5){\line(1,0){30}}
\put(217,55.5){\mbox{$\times$}}
\put(205,-2){\circle*{6}}\put(235,-2){\circle*{6}}
\qbezier(198,-2)(198,-9)(205,-9)\put(204,-9){\vector(1,0){2}}
\qbezier(242,-2)(242,-9)(235,-9)\put(235,-9){\vector(-1,0){2}}
\put(205,-2.5){\line(1,0){30}}\put(205,-1.5){\line(1,0){30}}
\put(217,-4.5){\mbox{$\times$}}
 }

\put(-35,0){
\put(205,58){\circle*{6}}\put(235,58){\circle*{6}}
}

\put(115,-15){
\put(205,58){\circle*{6}}\put(235,58){\circle*{6}}
}

\put(-35,-60){
\put(205,58){\circle*{6}}\put(235,58){\circle*{6}}
}

\put(185,-15){
\put(205,58){\circle*{6}}\put(235,58){\circle*{6}}
\qbezier(198,58)(198,51)(205,51)\put(204,51){\vector(1,0){2}}
\qbezier(242,58)(242,51)(235,51)\put(235,51){\vector(-1,0){2}}
\put(205,57.5){\line(1,0){30}}\put(205,58.5){\line(1,0){30}}
\put(217,55.5){\mbox{$\times$}}
}

\put(255,-15){
\put(205,58){\circle*{6}}\put(235,58){\circle*{6}}
\qbezier(198,58)(198,51)(205,51)\put(204,51){\vector(1,0){2}}
\qbezier(242,58)(242,51)(235,51)\put(235,51){\vector(-1,0){2}}
\put(205,57.5){\line(1,0){30}}\put(205,58.5){\line(1,0){30}}
\put(217,55.5){\mbox{$\times$}}
}

 \put(185,-5){
\put(340,48){\circle*{6}}\put(380,48){\circle*{6}}
\qbezier(333,48)(333,41)(340,41)\put(339,41){\vector(1,0){2}}
\qbezier(387,48)(387,41)(380,41)\put(380,41){\vector(-1,0){2}}
\qbezier(340,47,5)(360,67,5)(380,47,5)
\qbezier(340,48,5)(360,69,5)(380,48,5)
\qbezier(340,47,5)(360,27,5)(380,47,5)
\qbezier(340,46,5)(360,25,5)(380,46,5)
\put(356.5,55.5){\mbox{$\times$}}\put(356.5,34.5){\mbox{$\times$}}
 }
\put(95,-82){\circle*{4}} \put(95,-88){\circle*{4}}
\put(355,-82){\circle{4}} \put(355,-88){\circle{4}}
\put(225,-60){\circle{4}}\put(225,-66){\circle*{4}}
\put(225,-103){\circle*{4}}\put(225,-109){\circle{4}}
\put(105,-85){\line(4,1){120}}\put(105,-85){\line(4,-1){120}}
\put(345,-85){\line(-4,1){120}}\put(345,-85){\line(-4,-1){120}}
\put(170,-87){\mbox{${\cal H}^{\cal L}$}}
\put(200,70){\mbox{$\Pi^{ik}_{ik} = N^2-N=N(N-1)$}}
\put(377,70){\mbox{$\Pi^{ik}_{jl}\Pi^{jl}_{ik} = N^2-2\cdot N+N^2 = 2N(N-1)$}}
\end{picture}

\noindent
Quantization is again straightforward:
\be
d^{\rm Hopf}_{S} = d^{\rm Hopf}_{\bullet\bullet} = [N]^2, \ \ \ \ \ \ \
d^{\rm Hopf}_{\bullet\circ} = d^{\rm Hopf}_{\circ\bullet} = [N][N-1], \ \ \ \ \ \ \
d^{\rm Hopf}_{\circ\circ} = [2][N][N-1]
\ee
and HOMFLY polynomials for four different initial vertices are:
\be
\!\!\!\!\!\!\!\!\!\!\!\!\!\!\!\!\!\!
q^{2(N-1)}\Big(d^{\rm Hopf}_{\bullet\bullet} - q(d^{\rm Hopf}_{\bullet\circ}
+d^{\rm Hopf}_{\circ\bullet}) + q^2\cdot d^{\rm Hopf}_{\circ\circ}\Big)
= q^{2N-2}[N]\Big([N]-2q[N-1]+q^2[2][N-1]\Big) = q^{2N-2}[N]\Big([N]+q^{N+1}-q^{3-N}\Big), \nn \\
\!\!\!\!\!\!\!\!\!\!\!
q^{N-1}\cdot(-q^{-N})\Big(d^{\rm Hopf}_{\bullet\circ} - q\cdot d^{\rm Hopf}_{\bullet\bullet}
- q\cdot d^{\rm Hopf}_{\circ\circ} + q^2\cdot d^{\rm Hopf}_{\circ\bullet}\Big)
= -q^{-1}[N]\Big((1+q^2)[N-1]-q[N]-q[2][N-1]\Big) = [N]^2,
\ \ \ \ \ \ \nn \\
\!\!\!\!\!\!\!\!\!\!\!
q^{N-1}\cdot(-q^{-N})\Big(d^{\rm Hopf}_{\circ\bullet} - q\cdot d^{\rm Hopf}_{\bullet\bullet}
- q\cdot d^{\rm Hopf}_{\circ\circ} + q^2\cdot d^{\rm Hopf}_{\bullet\circ}\Big)
= -q^{-1}[N]\Big((1+q^2)[N-1]-q[N]-q[2][N-1]\Big) = [N]^2,
\ \ \ \ \ \ \nn \\
\!\!\!\!\!\!\!\!\!\!\!\!\!
q^{-2N}\Big(d^{\rm Hopf}_{\circ\circ} - q(d^{\rm Hopf}_{\bullet\circ}
+d^{\rm Hopf}_{\circ\bullet}) + q^2\cdot d^{\rm Hopf}_{\bullet\bullet}\Big)
= q^{-2N}[N]\Big([2][N-1]-2q[N-1]+q^2[N]\Big) = q^{2-2N}[N]\Big([N]+q^{-N-1}-q^{N-3}\Big)\nn
\ee
Two of them describe a pair of disconnected unknots, the other two, related by the change
$q\longrightarrow q^{-1}$, are the fundamental HOMFLY for the Hopf link {\it per se}.

\bigskip

{\bf Ex.4.} Another example with two-dimensional hypercube is provided by
\underline{virtual trefoil}

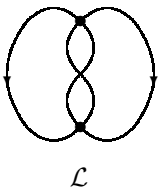
\begin{picture}(300,100)(-200,-40)
%
%
  \put(-80,0){
    \put(0,-10){\circle*{4}}\put(0,30){\circle*{4}}
    \linethickness{0.2mm}
    \qbezier(0,-10)(10,0)(0,10) \qbezier(0,-10)(-10,0)(0,10)
    \put(0,20){\qbezier(0,-10)(10,0)(0,10) \qbezier(0,-10)(-10,0)(0,10)}
    \qbezier(0,-10)(10,-20)(20,-10)
    \put(-20,0){\qbezier(0,-10)(10,-20)(20,-10)}
    \put(0,30){\qbezier(0,0)(10,10)(20,0)}
    \put(-20,30){\qbezier(0,0)(10,10)(20,0)}
    \put(-20,-10){\qbezier(0,0)(-15,20)(0,40)}
    \put(20,-10){\qbezier(0,0)(15,20)(0,40)}
    \put(-27.5,7){\vector(0,-1){2}}\put(27.5,7){\vector(0,-1){2}}
  }
\put(-84,-32){\mbox{${\cal L}$}}
\end{picture}

\newpage
\noindent
we have two real intersections and one virtual, thus negative dimensions do again appear:

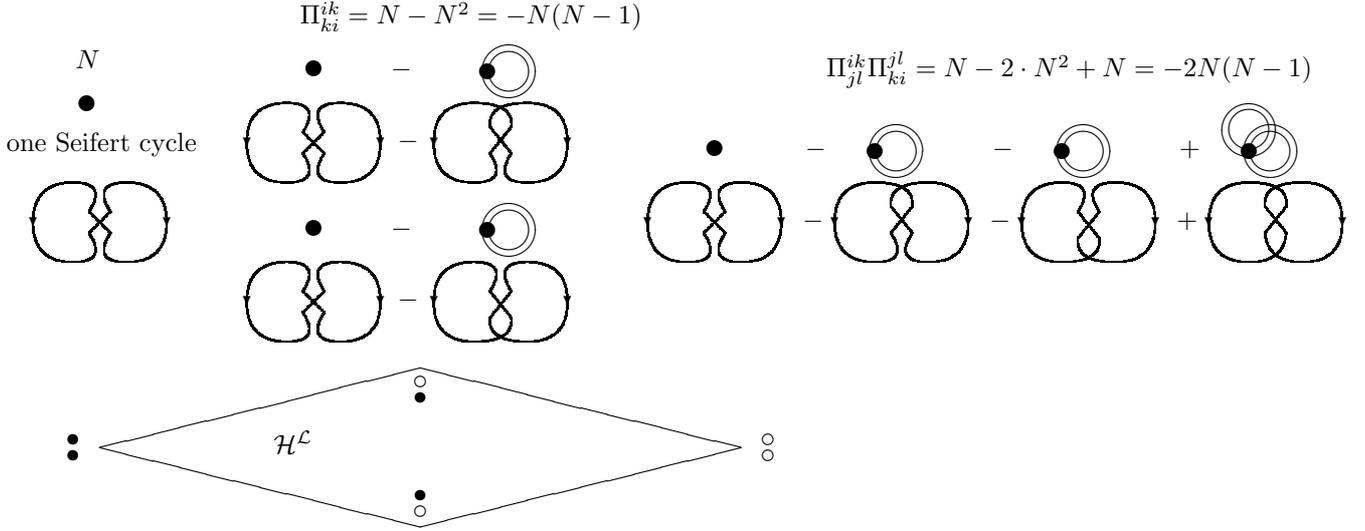
\begin{picture}(300,220)(90,-125)
  %
  %
  \linethickness{0.2mm}
\qbezier(80,0)(80,15)(95,15)\qbezier(80,0)(80,-15)(95,-15)
\qbezier(101,4)(108,15)(95,15)\qbezier(95,-15)(108,-15)(101,-4)
\qbezier(101,4)(105,0)(109,-4) \qbezier(109,4)(105,0)(101,-4)
\qbezier(109,4)(102,15)(115,15)\qbezier(115,-15)(102,-15)(109,-4)
\qbezier(130,0)(130,15)(115,15)\qbezier(130,0)(130,-15)(115,-15)
\put(80,0){\vector(0,-1){2}}\put(130,0){\vector(0,-1){2}}
%
%
%
\qbezier(160,30)(160,45)(175,45)\qbezier(160,30)(160,15)(175,15)
\qbezier(181,34)(188,45)(175,45)\qbezier(175,15)(188,15)(181,26)
\qbezier(181,34)(185,30)(189,26)\qbezier(189,34)(185,30)(181,26)
\qbezier(189,34)(182,45)(195,45)\qbezier(195,15)(182,15)(189,26)
\qbezier(210,30)(210,45)(195,45)\qbezier(210,30)(210,15)(195,15)
\put(160,30){\vector(0,-1){2}}\put(210,30){\vector(0,-1){2}}
\qbezier(230,30)(230,45)(245,45)\qbezier(230,30)(230,15)(245,15)
\qbezier(259,36)(260,45)(245,45)\qbezier(251,36)(250,45)(265,45)
\qbezier(259,36)(255,31)(251,26) \qbezier(251,36)(255,31)(259,26)
\qbezier(245,15)(258,15)(251,26)\qbezier(265,15)(252,15)(259,26)
\qbezier(280,30)(280,45)(265,45)\qbezier(280,30)(280,15)(265,15)
\put(230,30){\vector(0,-1){2}}\put(280,30){\vector(0,-1){2}}
\put(216.5,28){\mbox{$-$}}
%
%
%
\qbezier(160,-30)(160,-45)(175,-45)\qbezier(160,-30)(160,-15)(175,-15)
\qbezier(181,-34)(188,-45)(175,-45)\qbezier(175,-15)(188,-15)(181,-26)
\qbezier(181,-34)(185,-30)(189,-26)\qbezier(189,-34)(185,-30)(181,-26)
\qbezier(189,-34)(182,-45)(195,-45)\qbezier(195,-15)(182,-15)(189,-26)
\qbezier(210,-30)(210,-45)(195,-45)\qbezier(210,-30)(210,-15)(195,-15)
\put(160,-30){\vector(0,-1){2}}\put(210,-30){\vector(0,-1){2}}
\qbezier(230,-30)(230,-45)(245,-45)\qbezier(230,-30)(230,-15)(245,-15)
\qbezier(245,-15)(258,-15)(251,-26)\qbezier(265,-15)(252,-15)(259,-26)
\qbezier(259,-36)(260,-45)(245,-45)\qbezier(251,-36)(250,-45)(265,-45)
\qbezier(259,-36)(255,-31)(251,-26) \qbezier(251,-36)(255,-31)(259,-26)
\qbezier(280,-30)(280,-45)(265,-45)\qbezier(280,-30)(280,-15)(265,-15)
\put(230,-30){\vector(0,-1){2}}\put(280,-30){\vector(0,-1){2}}
\put(216.5,-32){\mbox{$-$}}
%
%
%
\qbezier(310,0)(310,15)(325,15)\qbezier(310,0)(310,-15)(325,-15)
\qbezier(331,4)(338,15)(325,15)\qbezier(325,-15)(338,-15)(331,-4)
\qbezier(331,4)(335,0)(339,-4)\qbezier(339,4)(335,0)(331,-4)
\qbezier(339,4)(332,15)(345,15)\qbezier(345,-15)(332,-15)(339,-4)
\qbezier(360,0)(360,15)(345,15)\qbezier(360,0)(360,-15)(345,-15)
\put(310,0){\vector(0,-1){2}}\put(360,0){\vector(0,-1){2}}
\put(368,-2){\mbox{$-$}}
\put(150,-30){
  \qbezier(230,30)(230,45)(245,45)\qbezier(230,30)(230,15)(245,15)
  \qbezier(259,36)(260,45)(245,45)\qbezier(251,36)(250,45)(265,45)
  \qbezier(259,36)(255,31)(251,26) \qbezier(251,36)(255,31)(259,26)
  \qbezier(245,15)(258,15)(251,26)\qbezier(265,15)(252,15)(259,26)
  \qbezier(280,30)(280,45)(265,45)\qbezier(280,30)(280,15)(265,15)
  \put(230,30){\vector(0,-1){2}}\put(280,30){\vector(0,-1){2}}
}
\put(438,-2){\mbox{$-$}}
\put(220,30){
  \qbezier(230,-30)(230,-45)(245,-45)\qbezier(230,-30)(230,-15)(245,-15)
  \qbezier(245,-15)(258,-15)(251,-26)\qbezier(265,-15)(252,-15)(259,-26)
  \qbezier(259,-36)(260,-45)(245,-45)\qbezier(251,-36)(250,-45)(265,-45)
  \qbezier(259,-36)(255,-31)(251,-26) \qbezier(251,-36)(255,-31)(259,-26)
  \qbezier(280,-30)(280,-45)(265,-45)\qbezier(280,-30)(280,-15)(265,-15)
  \put(230,-30){\vector(0,-1){2}}\put(280,-30){\vector(0,-1){2}}
}
\put(508,-2){\mbox{$+$}}
\put(290,-30){
  \qbezier(230,30)(230,45)(245,45)\qbezier(230,30)(230,15)(245,15)
  \qbezier(259,36)(260,45)(245,45)\qbezier(251,36)(250,45)(265,45)
  \qbezier(259,36)(255,31)(251,26) \qbezier(251,36)(255,31)(259,26)
  \qbezier(245,15)(258,15)(259,26)\qbezier(265,15)(252,15)(251,26)
  \qbezier(280,30)(280,45)(265,45)\qbezier(280,30)(280,15)(265,15)
  \put(230,30){\vector(0,-1){2}}\put(280,30){\vector(0,-1){2}}
}
%
%
\put(100,45){\circle*{6}} 
\put(70,27){\mbox{one Seifert cycle}}
\put(96,58){\mbox{$N$}} 
%
%
%
\put(10,10){
  \put(75,3){\put(100,45){\circle*{6}}}
  \put(200,5){
    \put(4,40){\mbox{$-$}}
    \put(40,42){\circle*{6}}
    \put(48,42){\circle{15}}
    \put(48,42){\circle{18}}
  }
}
%
%
\put(10,-50){
  \put(75,3){\put(100,45){\circle*{6}}}
  \put(200,5){
    \put(4,40){\mbox{$-$}}
    \put(40,42){\circle*{6}}
    \put(48,42){\circle{15}}
    \put(48,42){\circle{18}}
  }
}
%
%
\put(250,-20){
  \put(-15,3){\put(100,45){\circle*{6}}}
  \put(115,5){
    \put(4,40){\mbox{$-$}}
    \put(30,42){\circle*{6}}
    \put(38,42){\circle{15}}
    \put(38,42){\circle{18}}
  }
  \put(185,5){
    \put(4,40){\mbox{$-$}}
    \put(30,42){\circle*{6}}
    \put(38,42){\circle{15}}
    \put(38,42){\circle{18}}
  }
  \put(255,5){
    \put(4,40){\mbox{$+$}}
    \put(30,42){
      \put(0,0){\circle*{6}}
      \put(8,0){\circle{15}}
      \put(8,0){\circle{18}}
      \put(0,8){\circle{15}}
      \put(0,8){\circle{18}}
    }
  }
}
%
%
%
\put(95,-82){\circle*{4}} \put(95,-88){\circle*{4}}
\put(355,-82){\circle{4}} \put(355,-88){\circle{4}}
\put(225,-60){\circle{4}}\put(225,-66){\circle*{4}}
\put(225,-103){\circle*{4}}\put(225,-109){\circle{4}}
\put(105,-85){\line(4,1){120}}\put(105,-85){\line(4,-1){120}}
\put(345,-85){\line(-4,1){120}}\put(345,-85){\line(-4,-1){120}}
\put(170,-87){\mbox{${\cal H}^{\cal L}$}}
%
\put(180,75){\mbox{$\Pi^{ik}_{ki} = N-N^2=-N(N-1)$}}
\put(377,55){\mbox{$\Pi^{ik}_{jl}\Pi^{jl}_{ki} = N-2\cdot N^2+N = -2N(N-1)$}}
\end{picture}

Quantization gives:
\be
d^{\rm vT}_{S} = d^{\rm vT}_{\bullet\bullet} = [N], \ \ \ \ \ \ \
d^{\rm vT}_{\bullet\circ} = d^{\rm vT}_{\circ\bullet} = - [N][N-1], \ \ \ \ \ \ \
d^{\rm vT}_{\circ\circ} = - [2][N][N-1]
\ee
and HOMFLY polynomials for the four different initial vertices are:
\be
\!\!\!\!\!\!\!\!\!\!\!\!\!\!\!\!\!\!
q^{2(N-1)}\Big(d^{\rm vT}_{\bullet\bullet} - q(d^{\rm vT}_{\bullet\circ}
+d^{\rm vT}_{\circ\bullet}) + q^2\cdot d^{\rm vT}_{\circ\circ}\Big)
= q^{2N-2}[N]\Big(1+2q[N-1]-q^2[2][N-1]\Big) = q^{2N-2}[N]\Big(1-q^{N+1}+q^{3-N}\Big), \nn \\
\!\!\!\!\!\!\!\!\!\!\!
q^{N-1}\cdot(-q^{-N})\Big(d^{\rm vT}_{\bullet\circ} - q\cdot d^{\rm vT}_{\bullet\bullet}
- q\cdot d^{\rm vT}_{\circ\circ} + q^2\cdot d^{\rm vT}_{\circ\bullet}\Big)
= -q^{-1}[N]\Big(-(1+q^2)[N-1]-q \cdot 1+q[2][N-1]\Big) = [N],
\ \ \ \ \ \ \nn \\
\!\!\!\!\!\!\!\!\!\!\!
q^{N-1}\cdot(-q^{-N})\Big(d^{\rm vT}_{\circ\bullet} - q\cdot d^{\rm vT}_{\bullet\bullet}
- q\cdot d^{\rm vT}_{\circ\circ} + q^2\cdot d^{\rm vT}_{\bullet\circ}\Big)
= -q^{-1}[N]\Big(-(1+q^2)[N-1]-q \cdot 1+q[2][N-1]\Big) = [N],
\ \ \ \ \ \ \nn \\
\!\!\!\!\!\!\!\!\!\!\!\!\!
q^{-2N}\Big(d^{\rm vT}_{\circ\circ} - q(d^{\rm vT}_{\bullet\circ}
+d^{\rm vT}_{\circ\bullet}) + q^2\cdot d^{\rm vT}_{\bullet\bullet}\Big)
= q^{-2N}[N]\Big(-[2][N-1]+2q[N-1]+q^2\cdot 1\Big) = q^{2-2N}[N]\Big(1-q^{-N-1}+q^{N-3}\Big)\nn
\ee

Two of them ($\bullet \circ$ and $\circ \bullet$) describe virtual unknot,
which is the same as the usual unknot.
The other two ($\bullet \bullet$ and $\circ \circ$) describe virtual trefoil
and its image under mirror reflection.

\section{Matrix model description at $q=1$}

\subsection{Dimensions from fat graphs}

At $q=1$ all dimensions of the secondary hypercube can be obtained
from matrix model considerations.
Namely, each dimension is associated with the ribbon graph
and is equal to particular Feynman diagram in a Gaussian matrix model.
The matrix model propagator is given by
\be
  \Pi^{ik}_{jl} = \ <M^i_j\, M^k_l>\ = \delta^i_j \delta^k_l - \delta^i_l \delta^j_k
\label{prop}
\ee
It is not invertible, therefore the would-be action of the matrix model is singular,
but the action is not really needed in most matrix-model considerations.

One thing about this description is a little unusual.
Namely, instead of a correlator of some operators $\ \Tr M^n\ $
(which is a natural object from matrix model point of view)
we associate with ${\cal L}$ "projection" of this correlator --
only selected Feynman diagrams from the set, contributing to the correlator,
enter expression for the knot polynomials.
This diagram-selection is a somewhat strange procedure and one can hope that its better
interpretation and treatment will be found in the future.

Anyhow, what one needs is the following.
From original $(2,2)$-valent graph ${\cal L}$ we build an absolutely different
fat graph $\Gamma^{\cal L}$.
Its potentially multi-valent vertices are in one-to-one correspondence with the Seifert cycles
of ${\cal L}$ and edges are associated with the changes of resolutions (colorings) of ${\cal L}$.
In other words, the set of vertices of $\Gamma^{\cal L}$ is associated with the
{\it base} (Seifert) vertex $v_S \in {\cal H}^{\cal L}$, while the full $\Gamma^{\cal L}$
is associated with the anti-Seifert vertex $v_{\bar S}\in {\cal H}^{\cal L}$.
At other vertices $v$ of the hypercube, lying in between $v_S$ and $v_{\bar S}$,
we get certain (not arbitrary)
fat subgraphs $\gamma_v \subset \Gamma^{\cal L}$
with some edges removed.
Classical dimension $d_v$ is given by the Feynman diagram, associated with $\gamma_v$,
where edges are substituted by propagators convoluted at the vertices.
The number of edges in $\Gamma^{\cal L}$ is the same as in the hypercube ${\cal H}^{\cal L}$,
but $\Gamma^{\cal L}\neq {\cal H}^{\cal L}$ -- even the number of vertices is totally
different.

In above examples we have:
\be
{\bf Ex.1:} & <\Tr I\ \Tr I>\ = N\cdot N = N^2, &
<:\Tr M:\ :\Tr M:>\ = \Pi^{ik}_{ik} = N^2-N = N(N-1), \nn \\ \nn \\
{\bf Ex.2:} & <\Tr I>\ = N, &   <\Tr M^2>\ = \Pi^{ik}_{ki} = N-N^2=-N(N-1), \nn \\ \nn \\
{\bf Ex.3:} & <\Tr I\ \Tr I>\ = N^2, & <:\Tr M:\ :\Tr M:>\ = N(N-1), \nn \\
&&
<:\Tr M^2:\ :\Tr M^2:>\ = \Pi^{ik}_{jl}\Pi^{jl}_{ik} = 2N(N-1), \nn \\ \nn \\
{\bf Ex.4:} & <\Tr I>\ = N, & <\Tr M^2>\ = \Pi^{ik}_{ki} = N-N^2=-N(N-1), \nn \\ \nn \\
&& <\Tr M\ M\ M\ M>= \Pi^{ik}_{jl}\Pi^{jl}_{ki} = -2N(N-1)
\ee
\begin{picture}(0,0)
  \put(253,30){\line(0,1){5}}
  \put(253,35){\line(1,0){30}\line(0,-1){5}}
  \put(12,0){
    \put(253,30){\line(0,1){10}}
    \put(253,40){\line(1,0){30}\line(0,-1){10}}
  }
\end{picture}

In first three examples projection to  $\Gamma^{\cal L}$ and even to particular
$\gamma_v\subset \Gamma^{\cal L}$ can be described by adjustment of operators
and normal orderings, but in general this does not work, as seen already in example 4:
one should just pick up particular Feynman diagram from particular correlator.
Still what one has is the clear interpretation of fat graphs $\gamma$
as Feynman diagrams, such that {\bf classical dimensions $d_{\gamma}$ are
the values of diagrams for propagator (\ref{prop})}.

To summarize, with a fat graph $\gamma$ we can associate a matrix-model correlator
$d_\gamma^{P}$ with the propagator  (\ref{prop}).

\subsection{Peculiarities}

Still, since the situation is somewhat unusual, we provide additional illustrations.

Consider, say, the average $<\tr M^4>$.
If it was the ordinary Gaussian matrix model (GMM)   with the propagator
$\ <M^i_j\, M^k_l>_{_{\text{GMM}}} \ = \  \delta^i_l \delta^j_k$,
the answer contains three terms:
\be
<\tr M^4>_{_{\text{GMM}}} =2N^3 + N
\ee
the first two associated with planar and the third -- with non-planar fat graphs:

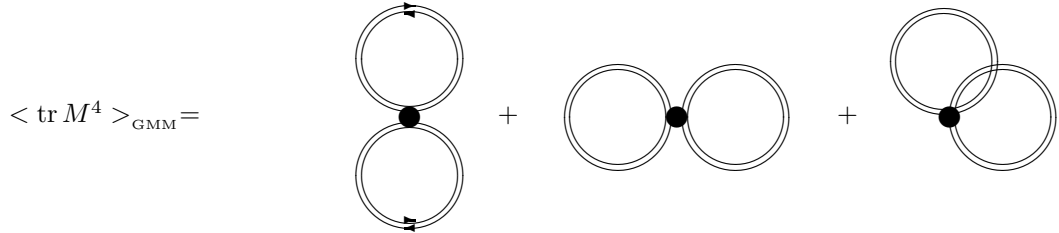
\begin{picture}(200,110)(-150,-50)
\put(-150,-2){\mbox{$<\tr M^4>_{_{\text{GMM}}} =$}}
\put(0,0){\circle*{8}}
\put(0,22){\circle{40}}
\put(0,22){\circle{34}}
\put(0,-22){\circle{34}}
\put(0,-22){\circle{40}}
\put(0,42){\vector(1,0){2}}
\put(0,39){\vector(-1,0){2}}
\put(0,-39){\vector(1,0){2}}
\put(0,-42){\vector(-1,0){2}}
\put(33,-2){\mbox{$+$}}
\put(100,0){\circle*{8}}
\put(78,0){\circle{40}}
\put(78,0){\circle{36}}
\put(122,0){\circle{36}}
\put(122,0){\circle{40}}
\put(160,-2){\mbox{$+$}}
\put(202,0){\circle*{8}}
\put(200,21){\circle{40}}
\put(200,21){\circle{36}}
\put(222,0){\circle{36}}
\put(222,0){\circle{40}}
\end{picture}

\bigskip


\noindent
We put arrows only in one place to avoid overloading the pictures.

\bigskip

$\bullet$ The first thing which happens when we substitute the propagator by (\ref{claprop})
to get the knot fat graphs (KFG)
is that each of these three diagrams turn into four:

\begin{picture}(200,335)(-150,-250)
\put(-120,28){\mbox{$<\tr M^4>_{_{\text{KFG}}} \ \ \ =$}}
\put(0,30){\circle*{8}}
\put(0,52){\circle{40}}
\put(0,52){\circle{36}}
\put(0,8){\circle{36}}
\put(0,8){\circle{40}}
\put(33,28){\mbox{$+$}}
\put(100,30){\circle*{8}}
\put(78,30){\circle{40}}
\put(78,30){\circle{36}}
\put(122,30){\circle{36}}
\put(122,30){\circle{40}}
\put(160,28){\mbox{$+$}}
\put(202,30){\circle*{8}}
\put(200,51){\circle{40}}
\put(200,51){\circle{36}}
\put(222,30){\circle{36}}
\put(222,30){\circle{40}}
\put(-40,-52){\mbox{$-$}}
\put(0,-50){\circle*{8}}
\put(0,-72){\circle{36}}
\put(0,-72){\circle{40}}
\put(33,-52){\mbox{$-$}}
\put(100,-50){\circle*{8}}
\put(122,-50){\circle{36}}
\put(122,-50){\circle{40}}
\put(160,-52){\mbox{$-$}}
\put(202,-50){\circle*{8}}
\put(222,-50){\circle{36}}
\put(222,-50){\circle{40}}
\put(-40,-152){\mbox{$-$}}
\put(0,-150){\circle*{8}}
\put(0,-128){\circle{40}}
\put(0,-128){\circle{36}}
\put(33,-152){\mbox{$-$}}
\put(100,-150){\circle*{8}}
\put(78,-150){\circle{40}}
\put(78,-150){\circle{36}}
\put(160,-152){\mbox{$-$}}
\put(202,-150){\circle*{8}}
\put(200,-129){\circle{40}}
\put(200,-129){\circle{36}}
%
\put(-40,-202){\mbox{$+$}}
\put(0,-200){\circle*{8}}
\put(33,-202){\mbox{$+$}}
\put(100,-200){\circle*{8}}
\put(160,-202){\mbox{$+$}}
\put(202,-200){\circle*{8}}
%
\put(-70,-235){\mbox{$= \ \ \ 2\cdot (N^3-2N^2+N)+(N-2N^2+N)\ = \ 2\cdot N(N-1)^2 -2N(N-1)$}}
\end{picture}

\bigskip

$\bullet$ The second thing is that different columns, all contributing to a single matrix-model average,
are associated with different knots: any of the first two -- with a triple of unknots (see below),
while the third one -- with the virtual trefoil.

Actually, given particular ribbon graph, we can algorithmically obtain a
planar diagram of a knot or link, hypercube construction for which involves this ribbon graph,
 though generally this planar diagram will be {\it virtual}
and what we get will not be the simplest possible diagram for this ribbon graph.

Let's illustrate this with the third diagram in the above example of $\ <\Tr M^4>$.
First we draw the fat graph on the plane in some way

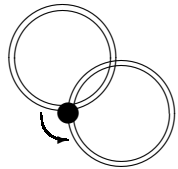
\begin{picture}(200,70)(-220,-25)
\put(2,0){
  \put(0,0){\circle*{8}}
  \linethickness{0.2mm} \qbezier(-10,0)(-10,-10)(0,-10) \put(0,-10){\vector(1,0){0}}
}

\put(0,21){\circle{40}}
\put(0,21){\circle{36}}
\put(22,0){\circle{36}}
\put(22,0){\circle{40}}
\end{picture}

\noindent
We explicitly show the orientation of the vertex, because now we think about it
as very tiny Seifert cycle, which already has orientation.

There are two problems, which stop us from converting this diagram into planar diagram of a knot
straight away. First, two edges of the projection of the fat graph intersect with each other.
Situations like this can be cured by the following transformation:

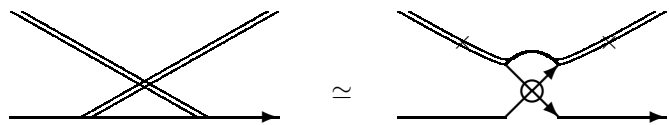
\begin{picture}(200,80)(-100,-20)
  \linethickness{0.5mm}
  \thicklines
  \qbezier(0,0)(50,0)(100,0)
  \put(100,0){\vector(1,0){0}}
      {\linethickness{0.1mm}
        \qbezier(0,40)(35,20)(70,0)
        \put(4,0){\qbezier(0,40)(35,20)(70,0)}
        \qbezier(100,40)(65,20)(30,0)
        \put(-4,0){\qbezier(100,40)(65,20)(30,0)}
  }
  \put(120,7){$\simeq$}
  \put(145,0){
    \linethickness{0.5mm}
    \thicklines
    \qbezier(0,0)(20,0)(40,0) \qbezier(60,0)(80,0)(100,0)
    \put(40,0){
      \thicklines
      \put(0,0){\vector(1,1){20}}
      \put(0,20){\vector(1,-1){20}}
      \put(10,10){\circle{8}}
    }
    \qbezier(40,20)(50,30)(60,20)
    \put(100,0){\vector(1,0){0}}
        {\linethickness{0.1mm}
          \qbezier(0,40)(35,20)(40,20)
          \put(4,0){\qbezier(0,40)(35,20)(37,22)}
          \put(20,26){$\times$}
          \qbezier(100,40)(65,20)(60,20)
          \put(-4,0){\qbezier(100,40)(65,20)(63,22)}
          \put(75,26){$\times$}
        }
  }
\end{picture}

\noindent
Here the piece of the Seifert cycle is drawn in bold, while the fat-graph edges are drawn as
thin double lines.

Note, that this transformation changes twisted edges to untwisted and vice versa.
Hence, after this transformation our example diagram becomes

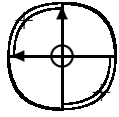
\begin{picture}(200,80)(-150,-40)
  \linethickness{0.4mm} \thicklines
  \qbezier(0,0)(20,0)(40,0) \qbezier(20,-20)(20,0)(20,20)
  \put(0,0){\vector(-1,0){0}} \put(20,20){\vector(0,1){0}}
  \put(20,0){\circle{8}}
  \qbezier(0,0)(0,-20)(20,-20)
  \qbezier(20,20)(40,20)(40,0)
  {\linethickness{0.15mm}
    \qbezier(0,0)(0,20)(20,20)
    \qbezier(2,0)(2,18)(20,18)
    \put(2,11){$+$}
  }
  {\linethickness{0.15mm}
    \qbezier(20,-20)(40,-20)(40,0)
    \qbezier(20,-18)(38,-18)(38,0)
    \put(31,-16){$+$}
  }


\end{picture}

And actually in this case this is it -- making the edges a little bit shorter we
can now transform them into black intersections, obtaining planar diagram of virtual trefoil.

However, it may also happen, that after all edge intersections had been eliminated,
untwisted edges still remain on the picture. In this case, we use transformation

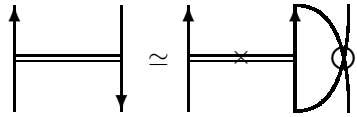
\begin{picture}(200,80)(-150,-20)
  \thicklines
  \put(0,0){\vector(0,1){40}}
  \put(40,40){\vector(0,-1){40}}
  {\thinlines \put(0,21){\line(1,0){40}} \put(0,19){\line(1,0){40}}}
  \put(50,17){$\simeq$}
  \put(65,0){
    \put(0,0){\vector(0,1){40}}
    \put(40,0){\vector(0,1){40}}
        {\thinlines \put(0,21){\line(1,0){40}} \put(0,19){\line(1,0){40}}}
        \put(16,18){$\times$}
        \qbezier(40,40)(60,40)(60,0) \qbezier(40,0)(60,0)(60,40)
        \put(58,20){\circle{8}}
  }
\end{picture}

\noindent
to convert them into twisted ones (which in turn are substituted by black intersections).

For example, in
the most elementary case we get:

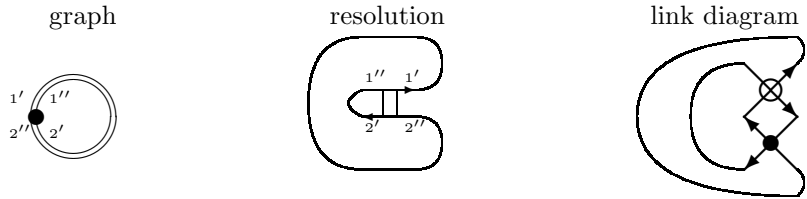
\begin{picture}(300,95)(-100,-40)
\put(5,35){\mbox{graph}}
\put(110,35){\mbox{resolution}}
\put(230,35){\mbox{link diagram}}
\put(0,0){\circle*{6}}
\put(14,0){\circle{30}}
\put(14,0){\circle{26}}
\put(-10,5){\mbox{{\tiny $1'$}}}
\put(5,5){\mbox{{\tiny $1''$}}}
\put(-10,-8){\mbox{{\tiny $2''$}}}
\put(5,-8){\mbox{{\tiny $2'$}}}
\put(80,0) {
  \put(50,0) {
    \put(0,0){\line(0,1){10}}
    \put(5,0){\line(0,1){10}}
    \put(3,4) {
      \put(-10,9){\mbox{{\tiny $1''$}}}
      \put(5,9){\mbox{{\tiny $1'$}}}
      \put(-10,-10){\mbox{{\tiny $2'$}}}
      \put(5,-10){\mbox{{\tiny $2''$}}}
    }
    \linethickness{0.2mm}
    \put(-8,0){\qbezier(0,0)(10,0)(20,0) \put(0,0){\vector(-1,0){0}}}
    \put(-8,10){\qbezier(0,0)(10,0)(20,0) \put(20,0){\vector(1,0){0}}}
    \qbezier(12,0)(22,0)(22,-10) \qbezier(22,-10)(22,-20)(12,-20)
    \put(0,30){\qbezier(12,0)(22,0)(22,-10) \qbezier(22,-10)(22,-20)(12,-20)}
    \put(12,-20){\qbezier(0,0)(-10,0)(-20,0)}
    \put(12,30){\qbezier(0,0)(-10,0)(-20,0)}
    \put(-8,0){\qbezier(0,0)(-10,5)(0,10)}
    \put(-8,-20){\qbezier(0,0)(-20,0)(-20,25) \qbezier(-20,25)(-20,50)(0,50)}
  }
}
%
%
\put(285,0){
  \thicklines
  \linethickness{0.2mm}
  \put(0,-20){\vector(-1,1){20}}
  \put(0,0){\vector(-1,-1){20}}
  \put(-10,-10){\circle*{6}}
  \put(-20,0){\vector(1,1){20}}
  \put(-20,20){\vector(1,-1){20}}
  \put(-10,10){\circle{8}}
  \qbezier(-20,-20)(-40,-20)(-40,0) \qbezier(-40,0)(-40,20)(-20,20)
  \qbezier(0,-20)(5,-25)(0,-30) \qbezier(0,-30)(-60,-30)(-60,0)
  \qbezier(-60,0)(-60,30)(0,30) \qbezier(0,30)(5,25)(0,20)
}
\end{picture}

\noindent
In other words the ordinary matrix-model graph, which has two closed loops and
therefore is equal to $N^2$, corresponds in our description to a pair of disconnected
circles. Note,however, that orientation matters: only for this choice we get one Seifert cycle
for the black resolution.
If mutual orientation is changed, then there are two Seifert cycles and {our} fat graph
has two vertices:

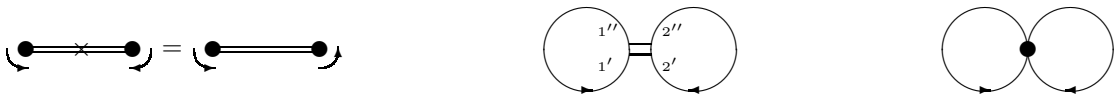
\begin{picture}(300,44)(-150,-15)
\put(-400,-48){
\put(290,48){\circle*{6}}\put(330,48){\circle*{6}}
\qbezier(283,48)(283,41)(290,41)\put(289,41){\vector(1,0){2}}
\qbezier(337,48)(337,41)(330,41)\put(331,41){\vector(-1,0){2}}
\put(290,47){\line(1,0){40}}\put(290,49){\line(1,0){40}}
\put(307,45.5){\mbox{$\times$}}
\put(341,46){\mbox{$=$}}
\put(360,48){\circle*{6}}\put(400,48){\circle*{6}}
\qbezier(353,48)(353,41)(360,41)\put(359,41){\vector(1,0){2}}
\qbezier(407,48)(407,41)(400,41)\put(407,47){\vector(0,1){2}}
\put(360,47){\line(1,0){40}}\put(360,49){\line(1,0){40}}
}
\put(100,0){\circle{30}} \put(100,-15.5){\vector(1,0){2}}
\put(140,0){\circle{30}}\put(140,-15.5){\vector(-1,0){2}}
\put(116,2){\line(1,0){8}}
\put(116,-2){\line(1,0){8}}
\put(104,5){\mbox{{\tiny $1''$}}}
\put(128,5){\mbox{{\tiny $2''$}}}
\put(104,-8){\mbox{{\tiny $1'$}}}
\put(128,-8){\mbox{{\tiny $2'$}}}
\put(50,0){
\put(199,0){\circle{30}}
\put(231,0){\circle{30}}
\put(215,0){\circle*{6}}
\put(199,-15.5){\vector(1,0){2}}
\put(231,-15.5){\vector(-1,0){2}}
}
\end{picture}

\bigskip

$\bullet$ The third thing is that association of link and Feynman diagrams is not
{\it quite} straightforward.
There is important difference between the standard  't Hooft's rule in matrix models,
when the end-point of a propagator is a double line with oppositely-directed arrows,
and the connection between Seifert cycles, where the trivial (black) resolution
corresponds to arrows pointing in the {\it same} direction.
This difference is the origin of twisted double lines in the examples  of section 1,
and, as already mentioned there,  this twisting is crucially important in our formalism.

To illustrate this difference, one and the same term $\delta^i_l\delta^k_j$ in the
propagator $\ <M^i_j,\ M^k_l>\ $ has different pictorial representation
in the ordinary matrix model (GMM) formalism  and in application to knots (KFG):

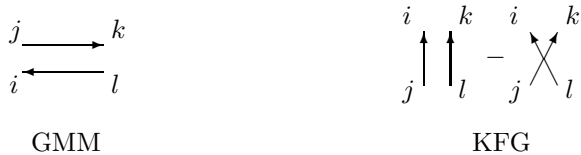
\begin{picture}(300,65)(-100,-25)
\put(0,10){\vector(1,0){30}}
\put(30,0){\vector(-1,0){30}}
\put(-5,13){\mbox{$j$}}
\put(33,13){\mbox{$k$}}
\put(-5,-8){\mbox{$i$}}
\put(33,-8){\mbox{$l$}}
\put(3,-30){\mbox{GMM}}
\put(150,-5){\vector(0,1){20}}
\put(160,-5){\vector(0,1){20}}
\put(142,-10){\mbox{$j$}}
\put(142,18){\mbox{$i$}}
\put(163,-10){\mbox{$l$}}
\put(163,18){\mbox{$k$}}
\put(173,3){\mbox{$-$}}
\put(190,-5){\vector(1,2){10}}
\put(200,-5){\vector(-1,2){10}}
\put(182,-10){\mbox{$j$}}
\put(182,18){\mbox{$i$}}
\put(203,-10){\mbox{$l$}}
\put(203,18){\mbox{$k$}}
\put(168,-30){\mbox{KFG}}
\end{picture}

\bigskip

\noindent
Accordingly, different are the "most natural" Feynman diagrams, say, in the following two cases:

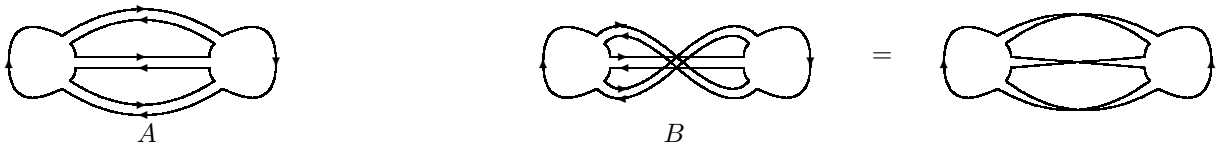
\begin{picture}(100,70)(0,-35)
\qbezier(0,0)(0,20)(20,10)
\qbezier(0,0)(0,-20)(20,-10)
\qbezier(100,0)(100,20)(80,10)
\qbezier(100,0)(100,-20)(80,-10)
\qbezier(20,10)(50,30)(80,10)
\qbezier(23,7)(50,25)(77,7)
\qbezier(23,7)(25,5)(25,2)  \qbezier(77,7)(75,5)(75,2)
\put(25,2){\line(1,0){50}}
\put(25,-2){\line(1,0){50}}
\qbezier(23,-7)(25,-5)(25,-2)   \qbezier(77,-7)(75,-5)(75,-2)
\qbezier(23,-7)(50,-25)(77,-7)
\qbezier(20,-10)(50,-30)(80,-10)
\put(50,20){\vector(1,0){2}}
\put(50,16){\vector(-1,0){2}}
\put(50,2){\vector(1,0){2}}
\put(50,-2){\vector(-1,0){2}}
\put(50,-16){\vector(1,0){2}}
\put(50,-20){\vector(-1,0){2}}
\put(0,0){\vector(0,1){2}}
\put(100,0){\vector(0,-1){2}}
\put(48,-30){\mbox{$A$}}
\qbezier(200,0)(200,20)(220,10)
\qbezier(200,0)(200,-20)(220,-10)
\qbezier(300,0)(300,20)(280,10)
\qbezier(300,0)(300,-20)(280,-10)
\qbezier(20,10)(50,30)(80,10)
\qbezier(23,7)(50,25)(77,7)
\qbezier(223,7)(225,5)(225,2)  \qbezier(277,7)(275,5)(275,2)
\qbezier(220,10)(230,20)(250,2)  \qbezier(250,2)(270,-16)(277,-7)
\qbezier(223,7)(230,16)(250,-2)  \qbezier(250,-2)(270,-20)(280,-10)
\put(225,2){\line(1,0){50}}
\put(225,-2){\line(1,0){50}}
\qbezier(220,-10)(230,-20)(250,-2)  \qbezier(250,-2)(270,16)(277,7)
\qbezier(223,-7)(230,-16)(250,2)  \qbezier(250,2)(270,20)(280,10)
\qbezier(223,-7)(225,-5)(225,-2)   \qbezier(277,-7)(275,-5)(275,-2)
\qbezier(23,-7)(50,-25)(77,-7)
\qbezier(20,-10)(50,-30)(80,-10)
\put(229,14){\vector(1,0){2}}
\put(230,10){\vector(-1,0){2}}
\put(230,2){\vector(1,0){2}}
\put(230,-2){\vector(-1,0){2}}
\put(230,-10){\vector(1,0){2}}
\put(229,-14){\vector(-1,0){2}}
\put(200,0){\vector(0,1){2}}
\put(300,0){\vector(0,-1){2}}
\put(245,-30){\mbox{$B$}}
\put(323,0){\mbox{$=$}}
\qbezier(350,0)(350,20)(370,10)
\qbezier(350,0)(350,-20)(370,-10)
\qbezier(450,0)(450,20)(430,10)
\qbezier(450,0)(450,-20)(430,-10)
\qbezier(370,10)(400,27.5)(427,7)
\qbezier(373,7)(400,27.5)(430,10)
\qbezier(373,7)(375,5)(375,2)  \qbezier(427,7)(425,5)(425,2)
\qbezier(375,2)(400,0)(425,-2)
\qbezier(375,-2)(400,0)(425,2)
\qbezier(373,-7)(375,-5)(375,-2)   \qbezier(427,-7)(425,-5)(425,-2)
\qbezier(373,-7)(400,-27.5)(430,-10)
\qbezier(370,-10)(400,-27.5)(427,-7)
%
%
\put(350,0){\vector(0,1){2}}
\put(450,0){\vector(0,1){2}}
\end{picture}

\bigskip

\noindent
$A$ and $B$ are the planar and non-planar contributions to the matrix-model correlator
$<:\Tr M^3: \ :\Tr M^3: >_{_{\text{GMM}}}$, and if redrawn in terms of twisted double-lines,
$B$ represents the trefoil.
As to $A$, from the knot-theory point of view it will correspond to the virtual Borommean rings
with three real and three virtual intersections.

\bigskip

$\bullet$ The fourth point concerns the notion of planarity.
In addition to the ordinary planarity of matrix-model diagrams,
a big role is now played by a kind of a complementary "cross-planarity".
As we just saw, ordinary knots are the extreme non-planar contributions from the
point of view of matrix-model formalism.
This is, of course, obvious, because the corresponding unreduced HOMFLY have the
lowest (first) degree in $N$.
Planar matrix-model diagrams instead have the maximal degree and describe multi-component
links -- generically, virtual.

However, if we express everything in terms of twisted (crossed) propagators -- i.e.
treat the {\it twisted} double line as {\it flat}, -- the situation becomes the opposite:
{\it cross-planar} diagrams represent knots, and the more is the degree of cross-non-planarity,
the bigger is the number of link components.

\subsection{Classical HOMFLY as dimension with a trivial (${\cal R}$-matrix) propagator}

Despite all these complications, at $q=1$ the direct outcome of matrix model formalism is
trivial.

An analogue $P^{\cal L}(T)$ of (\ref{qaltsum}) for $c=s$ (i.e. for the all-black coloring),
\be
P^{\cal L}(T) = \sum_{v \subset {\cal H}^{\cal L}} d_{\gamma_v}\cdot T^{E(\gamma_v)}
\label{TdefoH}
\ee
where $E(\gamma_v)=h_v-h_s$ is the number of edges in $\gamma_v$,
was named {\it primary $T$-deformation of HOMFLY} in \cite{AnoMKhR}.
Since amputation of a propagator is equivalent to changing it for $\delta^{i_j}\delta^{k_l}$,
this quantity can be considered as the correlator
\be
P^{\cal L}(T) = d_{\Gamma^{\cal L}}^{I+TP}
\ee
evaluated with the $T$-deformed (or rather $T^{-1}$-deformed) propagator
\be
\delta^i_j\delta^k_l  + T\cdot \Pi^{ik}_{jl}
\ee

Classical ($q=1$) HOMFLY polynomial arises at $T=-1$, when the propagator
$\delta^i_j\delta^k_l -\Pi^{ik}_{jl} =\delta^i_l\delta^k_j$,
becomes just the classical value of $R$-matrix (permutation).
Since such contraction of Seifert cycles provides just the original link,
we get
\be
H^{\cal L}(q=1,N) = P^{\cal L}(T=-1) = N^{\#\ \text{of link components}}
\ee

What makes things non-trivial is "quantization", i.e. switching on $q\neq 1$.
Relation of HOMFLY to ${\cal R}$-matrix propagator survives quantization,
but the  answer is no longer a power of $N$.
Quantization implies that the weight factors $(-)^s\longrightarrow (-q)^s$
in (\ref{TdefoH}), but also $d_\gamma$ change for $q$-dependent
quantum dimension -- thus providing a non-trivial quantization of shift operator.
Advantage of $P^{\cal L}(T)$ is that it can be easily quantized by the substitution
$\Pi \longrightarrow [2]\cdot\pi$,
where $\pi$ is  projector on the antisymmetric representation $[11]\subset [1]\otimes[1]$ --
and after that it plays a profound role in construction of Khovanov-Rozansky polynomials,
see \cite{AnoMKhR}).
For our purposes, this quantization of $P^{\cal L}(T)$ provides information about the quantum
dimensions $d_\gamma(q,N)$.

\subsection{Ward identities}

What makes matrix-model intuition really powerful \cite{UFN23} is the fact that
in field theory one can easily formulate quite general relations,
like
\be
< \Tr M \  :F(M): >\ = \Pi^{ik}_{il} \left<\frac{\p F}{\p M^k_l}\right>\  =
(N-1)\delta^k_l  \left<\frac{\p F}{\p M^k_l}\right>, \nn \\
< :\Tr M^2: \  :F(M): >\ = \Pi^{ik}_{jl}\Pi^{jm}_{in} \left<\frac{\p^2 F}{\p M^k_l\p M^m_n}\right>\  =
\Big((N-2)\delta^k_l\delta^m_n +\delta^k_n\delta^m_l\Big)
\left<\frac{\p^2 F}{\p M^k_l\p M^m_n}\right>, \nn \\
< :\Tr M^3: \  :F(M): >\ = \Pi^{il}_{jm}\Pi^{jn}_{kp}\Pi^{kr}_{is}
\left<\frac{\p^3 F}{\p M^l_m\p M^n_p\p M^r_s}\right>\  = \ \ \ \ \ \ \ \ \ \ \ \nn\\
\Big((N-3)\delta^l_m\delta^n_p\delta^r_s
+\left( \delta^n_m \delta^l_p \delta^r_s + \delta^l_m \delta^r_p \delta^n_s + \delta^n_p \delta^l_s \delta^r_m \right) - \delta^r_m \delta^l_p \delta^n_s \Big)
\left<\frac{\p^3 F}{\p M^l_m\p M^n_p\p M^r_s}\right>, \nn \\
\ldots \label{eq:q1-wards}
\ee
which in our context can be considered as certain recursion relations for classical dimensions.

Even more interesting, the first two of these relations are in fact related to
the first two Reidemeister moves and therefore survive quantization --
one should only change $N-1$ and $N-2$ for $[N-1]$ and $[N-2]$.
In other words, matrix model intuition allows us to convert Reidemeister invariance
into explicit recursion rules for quantum dimensions, which allows to calculate
them in a very effective way.
From practical point of view this provides for the hypercube approach an {\it analogue} of the
skein relation technique -- but now it does not rely on the properties of the fundamental
${\cal R}$-matrix and can be applied to more general case, like HOMFLY polynomials
for virtual knots.


\subsection{Tasting  $q \neq 1$}

At $q \neq 1$ most of recursion relations of the last section deform in some (yet)
uncontrollable way, to the point that one may wonder, whether it is at all possible
to calculate quantum dimensions in any other way than by explicitly computing
quantum traces of projectors, as done in \cite{AnoMKhR}.
In particular, naive attempts to quantize the $:\Tr M^3:$-elimination rule,
i.e. the third one in (\ref{eq:q1-wards}), and apply it to trefoil diagram
$\begin{picture}(30,10)(-5,-2)
  \put(0,0){
    \put(0,0){\circle*{6}}
    \put(0,0){\qbezier(-5,0)(-5,-5)(0,-5) \put(2,-5){\vector(1,0){0}}}
  }\put(20,0){
    \put(0,0){\circle*{6}}
    \put(0,0){\qbezier(5,0)(5,-5)(0,-5) \put(-2,-5){\vector(-1,0){0}}}
  }\qbezier(0,0)(10,10)(20,0)\qbezier(0,0)(10,0)(20,0)
\qbezier(0,0)(10,-10)(20,0)\end{picture}$) fail to produce the correct answer.

However, as already mentioned, there are recursion relations which survive, and we list them here
in the form of allowed (local) transformations of the ribbon graph
(i.e. quantum dimensions of lhs and rhs coincide).
We do not claim yet that this is exhaustive set of such preserved relations
--  these are just ones which are the easiest to find.

First line of \eqref{eq:q1-wards} becomes rule for elimination of 1-valent
vertices (\textbf{the $[N-1]$-rule})

\begin{picture}(300,60)(-130,-10)
  \linethickness{0.4mm} \thicklines
  \qbezier(0,0)(20,20)(0,40)
  \put(0,40){\vector(-1,1){0}}
  \put(35,10){\vector(1,0){0}}
  \put(30,20){\circle{20}}
      {\linethickness{0.15mm} \put(10,21){\line(1,0){10}} \put(10,19){\line(1,0){10}}}
      \put(50,17) {$\simeq\ \ [N-1]$}
      \put(110,0){\vector(0,1){40}}
\end{picture}

Next, there is a rule for elimination of double edges (\textbf{the $[2]$-rule})

\begin{picture}(300,60)(-130,-10)
  \linethickness{0.4mm} \thicklines
  \qbezier(0,0)(20,20)(0,40)
  \put(0,40){\vector(-1,1){0}}
  \put(40,0){\qbezier(0,0)(-20,20)(0,40)}
  \put(40,40){\vector(1,1){0}}
      {\linethickness{0.15mm}
        \put(8,31){\line(1,0){24}}
        \put(8,29){\line(1,0){24}}
        \put(8,11){\line(1,0){24}}
        \put(8,9){\line(1,0){24}}
        \put(16,28){$\times$}
        \put(16,8){$\times$}
      }
      \put(50,17) {$\simeq\ \ [2]$}
      \put(80,0){
        \linethickness{0.4mm} \thicklines
        \qbezier(0,0)(20,20)(0,40)
        \put(0,40){\vector(-1,1){0}}
        \put(40,0){\qbezier(0,0)(-20,20)(0,40)}
        \put(40,40){\vector(1,1){0}}
            {\linethickness{0.15mm}
              \put(10,21){\line(1,0){20}}
              \put(10,19){\line(1,0){20}}
              \put(16,18){$\times$}
            }
      }
\end{picture}

Second line of \eqref{eq:q1-wards} becomes rule for elimination of 2-valent
vertices (\textbf{the $[N-2]$-rule})

\begin{picture}(300,60)(-130,-10)
  \linethickness{0.4mm} \thicklines
  \put(-40,0){\qbezier(0,0)(20,20)(0,40) \put(0,40){\vector(-1,1){0}}}
  \put(40,0){\qbezier(0,0)(-20,20)(0,40)}
  \put(40,0){\vector(1,-1){0}}
  \put(-40,0){\linethickness{0.15mm}
    \put(10,21){\line(1,0){20}}
    \put(10,19){\line(1,0){20}}
    \put(16,18){$\times$}
  }
  \put(0,0){\linethickness{0.15mm}
    \put(10,21){\line(1,0){20}}
    \put(10,19){\line(1,0){20}}
    \put(16,18){$\times$}
  }
  \put(0,20){\circle{20} \put(5,10){\vector(1,0){0}} \put(-5,-10){\vector(-1,0){0}}}

      \put(50,17) {$\simeq\ \ [N-2]$}
      \put(100,0) {
        \linethickness{0.4mm} \thicklines
        \qbezier(0,0)(20,20)(0,40)
        \put(0,40){\vector(-1,1){0}}
        \put(40,0){\qbezier(0,0)(-20,20)(0,40)}
        \put(40,0){\vector(1,-1){0}}
      }
      \put(150,17){$+$}
      \put(165,0) {
        \linethickness{0.4mm} \thicklines
        \qbezier(0,0)(20,20)(40,0)
        \put(0,40){\vector(-1,1){0}}
        \put(40,40){\qbezier(0,0)(-20,-20)(-40,0)}
        \put(40,0){\vector(1,-1){0}}
      }
\end{picture}

Instead of third line of \eqref{eq:q1-wards} there is a ``mutation rule''
for 3-valent vertex (\textbf{the $[N-3]$-rule})

\begin{picture}(300,60)(-100,-10)
  \put(-40,0) {
    \linethickness{0.4mm} \thicklines
    \qbezier(0,0)(0,20)(-20,20) \put(-20,20){\vector(-1,0){0}}
    \put(20,0){\qbezier(0,0)(0,20)(20,20) \put(0,0){\vector(0,-1){0}}}
    \put(20,40){\qbezier(0,0)(-10,-10)(-20,0) \put(0,0){\vector(2,1){0}}}
    \put(10,20){\circle{16}}
    \put(10,0){
      \linethickness{0.15mm} \thinlines
      \qbezier(-12,10)(-12,10)(-5,15)
      \put(0,2){\qbezier(-12,10)(-12,10)(-7,14)}
      \put(-12,10){$+$}
    }
    \put(10,0){
      \linethickness{0.15mm} \thinlines
      \qbezier(12,10)(12,10)(5,15)
      \put(0,2){\qbezier(12,10)(12,10)(7,14)}
      \put(6,10){$+$}
    }
    \put(10,28){
      \linethickness{0.15mm} \thinlines
      \qbezier(-1,0)(-1,5)(-1,7)
      \qbezier(1,0)(1,5)(1,7)
      \put(-4,0){$\times$}
    }
  }
  \put(10,17){$-\ [N-3]$}
  \put(80,0) {
    \linethickness{0.4mm} \thicklines
    \qbezier(0,0)(0,20)(-20,20) \put(-20,20){\vector(-1,0){0}}
    \put(20,0){\qbezier(0,0)(0,20)(20,20) \put(0,0){\vector(0,-1){0}}}
    \put(20,40){\qbezier(0,0)(-10,-10)(-20,0) \put(0,0){\vector(2,1){0}}}
  }
  \put(130,17){$\simeq$}
  \put(210,0) {
    \put(-40,0) {
      \linethickness{0.4mm} \thicklines
      \qbezier(0,40)(0,20)(-20,20) \put(-20,20){\vector(-1,0){0}}
      \put(20,0){\qbezier(0,40)(0,20)(20,20) \put(0,40){\vector(0,1){0}}}
      \put(20,0){\qbezier(0,0)(-10,10)(-20,0) \put(0,0){\vector(2,-1){0}}}
      \put(10,20){\circle{16}}
      \put(10,40){
        \linethickness{0.15mm} \thinlines
        \qbezier(-12,-10)(-12,-10)(-5,-15)
        \put(0,-2){\qbezier(-12,-10)(-12,-10)(-7,-14)}
        \put(-12,-16){$+$}
      }
      \put(10,40){
        \linethickness{0.15mm} \thinlines
        \qbezier(12,-10)(12,-10)(5,-15)
        \put(0,-2){\qbezier(12,-10)(12,-10)(7,-14)}
        \put(6,-16){$+$}
      }
      \put(10,5){
        \linethickness{0.15mm} \thinlines
        \qbezier(-1,0)(-1,5)(-1,7)
        \qbezier(1,0)(1,5)(1,7)
        \put(-4,0){$\times$}
      }
    }
    \put(10,17){$-\ [N-3]$}
    \put(80,0) {
      \linethickness{0.4mm} \thicklines
      \qbezier(0,40)(0,20)(-20,20) \put(-20,20){\vector(-1,0){0}}
      \put(20,0){\qbezier(0,40)(0,20)(20,20) \put(0,40){\vector(0,1){0}}}
      \put(20,0){\qbezier(0,0)(-10,10)(-20,0) \put(0,0){\vector(2,-1){0}}}
    }
  }
\end{picture}

Also, there is 3-strand analogue of $[2]$-rule (\textbf{the $1$-rule}).
Again it is a ``mutation'' (it does not reduce complexity of the fat graph) not a ``decomposition''

\begin{picture}(300,60)(-100,-10)
  \linethickness{0.4mm} \thicklines
  \put(0,0) {
    \linethickness{0.4mm} \thicklines
    \qbezier(0,0)(0,20)(0,40)
    \put(0,40){\vector(0,1){0}}
    \put(-10,10){\linethickness{0.15mm}
      \put(10,21){\line(1,0){20}}
      \put(10,19){\line(1,0){20}}
      \put(16,18){$\times$}
    }
    \put(-10,-10){\linethickness{0.15mm}
      \put(10,21){\line(1,0){20}}
      \put(10,19){\line(1,0){20}}
      \put(16,18){$\times$}
    }
  }
  \put(20,0) {
    \linethickness{0.4mm} \thicklines
    \qbezier(0,0)(0,20)(0,40)
    \put(0,40){\vector(0,1){0}}
    \put(-10,0){\linethickness{0.15mm}
      \put(10,21){\line(1,0){20}}
      \put(10,19){\line(1,0){20}}
      \put(16,18){$\times$}
    }
  }
  \put(40,0) {
    \linethickness{0.4mm} \thicklines
    \qbezier(0,0)(0,20)(0,40)
    \put(0,40){\vector(0,1){0}}
  }
  \put(50,17){$-$}
  \put(65,0) {
    \linethickness{0.4mm} \thicklines
    \put(0,0) {
      \linethickness{0.4mm} \thicklines
      \qbezier(0,0)(0,20)(0,40)
      \put(0,40){\vector(0,1){0}}
      \put(-10,0){\linethickness{0.15mm}
        \put(10,21){\line(1,0){20}}
        \put(10,19){\line(1,0){20}}
        \put(16,18){$\times$}
      }
    }
    \put(20,0) {
      \linethickness{0.4mm} \thicklines
      \qbezier(0,0)(0,20)(0,40)
      \put(0,40){\vector(0,1){0}}
    }
    \put(40,0) {
      \linethickness{0.4mm} \thicklines
      \qbezier(0,0)(0,20)(0,40)
      \put(0,40){\vector(0,1){0}}
    }
  }
  \put(120,17){$\simeq$}
  \put(140,0) {
    \linethickness{0.4mm} \thicklines
    \put(0,0) {
      \linethickness{0.4mm} \thicklines
      \qbezier(0,0)(0,20)(0,40)
      \put(0,40){\vector(0,1){0}}
      \put(-10,0){\linethickness{0.15mm}
        \put(10,21){\line(1,0){20}}
        \put(10,19){\line(1,0){20}}
        \put(16,18){$\times$}
      }
    }
    \put(20,0) {
      \linethickness{0.4mm} \thicklines
      \qbezier(0,0)(0,20)(0,40)
      \put(0,40){\vector(0,1){0}}
      \put(-10,10){\linethickness{0.15mm}
        \put(10,21){\line(1,0){20}}
        \put(10,19){\line(1,0){20}}
        \put(16,18){$\times$}
      }
      \put(-10,-10){\linethickness{0.15mm}
        \put(10,21){\line(1,0){20}}
        \put(10,19){\line(1,0){20}}
        \put(16,18){$\times$}
      }
    }
    \put(40,0) {
      \linethickness{0.4mm} \thicklines
      \qbezier(0,0)(0,20)(0,40)
      \put(0,40){\vector(0,1){0}}
    }
    \put(50,17){$-$}
    \put(65,0) {
      \linethickness{0.4mm} \thicklines
      \put(0,0) {
        \linethickness{0.4mm} \thicklines
        \qbezier(0,0)(0,20)(0,40)
        \put(0,40){\vector(0,1){0}}
      }
      \put(20,0) {
        \linethickness{0.4mm} \thicklines
        \qbezier(0,0)(0,20)(0,40)
        \put(0,40){\vector(0,1){0}}
        \put(-10,0){\linethickness{0.15mm}
          \put(10,21){\line(1,0){20}}
          \put(10,19){\line(1,0){20}}
          \put(16,18){$\times$}
        }
      }
      \put(40,0) {
        \linethickness{0.4mm} \thicklines
        \qbezier(0,0)(0,20)(0,40)
        \put(0,40){\vector(0,1){0}}
      }
    }
  }
\end{picture}

Last, but not the least, dimensions behave nicely, when flips are
applied to the fat graph (\textbf{the flip-rule})

\begin{picture}(300,60)(-130,-10)
  \linethickness{0.4mm} \thicklines
  \put(0,0) {
    \linethickness{0.4mm} \thicklines
    \qbezier(0,0)(20,20)(0,40)
    \put(0,40){\vector(-1,1){0}}
    \put(40,0){\qbezier(0,0)(-20,20)(0,40)}
    \put(40,0){\vector(1,-1){0}}
    \put(0,0){\linethickness{0.15mm}
      \put(10,21){\line(1,0){20}}
      \put(10,19){\line(1,0){20}}
    }
  }
  \put(50,17){$\simeq\ \ -$}
  \put(75,0) {
    \linethickness{0.4mm} \thicklines
    \qbezier(0,0)(20,20)(40,0)
    \put(0,40){\vector(-1,1){0}}
    \put(40,40){\qbezier(0,0)(-20,-20)(-40,0)}
    \put(40,0){\vector(1,-1){0}}
    \put(0,0){\linethickness{0.15mm}
      \put(21,10){\line(0,1){20}}
      \put(19,10){\line(0,1){20}}
    }
  }
\end{picture}

The reason, why these rules are preserved by the quantization, is quite elegant.
Namely, $[N-1]$-rule is required to ensure invariance of result of the hypercube computation
 w.r.t 1st Reidemeister move (RM); $[2]$-rule is additionally needed to ensure invariance w.r.t
2nd RM with parallel strands; $[N-2]$-rule is needed for antiparallel 2nd RM;
 $[N-3]$- and $1$-rules ensure different orientations of 3rd RM.

In this nice explanation the flip-rule clearly stands out -- it is not required by Reidemeister
invariance, but we nevertheless observe, that it holds in all examples we calculate.
We can say that it is the most intriguing character in the whole story of $q \neq 1$.

\bigskip

While a lot of questions remain open, both conceptual and technical,
in practice these ideas lead to a serious enhancement of original
results of  \cite{DM3,AnoMKhR} and \cite{DM3virt}.
A program \cite{prog},
based on discovered recursions, easily reproduces fundamental
HOMFLY for entire Rolfsen table of \cite{katlas} (i.e. for all ordinary
knots with up to 10 intersections) -- but it re-deduces them from the hypercube
method.
As a far more spectacular and interesting illustration we now present
2-cabled HOMFLY for the simplest virtual knots -- this is a hypercube
calculation for 12 and 14 intersections, i.e. for hypercubes with up to
$2^{14}$ (!) vertices.

\section{On cabled virtual knots}

\subsection{2-cabled virtual trefoil -- first attempt}


Knot diagram of usual (non-cabled) virtual trefoil ($2.1$ in \cite{virtkatlas}) is

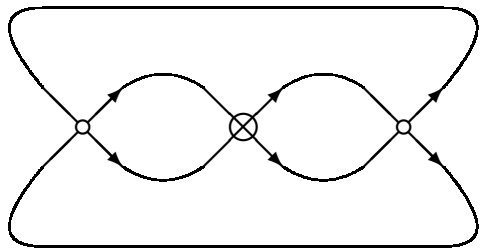
\begin{picture}(300,120)(-100,0)
  \put(100,50){
    \put(0,0){
      \thicklines
      \put(0,0){\vector(1,1){30}}
      \put(0,30){\vector(1,-1){30}}
      \put(15,15){\circle{10}}
    }
    \put(60,0){
      \thicklines
      \put(0,0){\line(1,1){13}}
      \put(17,17){\vector(1,1){13}}
      \put(0,30){\line(1,-1){13}}
      \put(17,13){\vector(1,-1){13}}
      \put(15,15){\circle{5}}
    }
    \put(-60,0){
      \thicklines
      \put(0,0){\line(1,1){13}}
      \put(17,17){\vector(1,1){13}}
      \put(0,30){\line(1,-1){13}}
      \put(17,13){\vector(1,-1){13}}
      \put(15,15){\circle{5}}
    }
    \put(30,30){\linethickness{0.2mm} \qbezier(0,0)(15,10)(30,0)}
    \put(30,0){\linethickness{0.2mm} \qbezier(0,0)(15,-10)(30,0)}
    \put(-30,30){\linethickness{0.2mm} \qbezier(0,0)(15,10)(30,0)}
    \put(-30,0){\linethickness{0.2mm} \qbezier(0,0)(15,-10)(30,0)}
    \put(-60,60){\thicklines \line(1,0){150}}
    \put(-60,-30){\thicklines \line(1,0){150}}
     \put(-60,-30){\linethickness{0.2mm} \qbezier(0,0)(-25,0)(0,30)}
    \put(-60,30){\linethickness{0.2mm} \qbezier(0,0)(-25,30)(0,30)}
    \put(90,-30){\linethickness{0.2mm} \qbezier(0,0)(25,0)(0,30)}
    \put(90,30){\linethickness{0.2mm} \qbezier(0,0)(25,30)(0,30)}
  }
\end{picture}

\noindent
and the corresponding HOMFLY polynomial was found  in \cite{DM3virt},
see also ex.4 in above sec.1:
\be
H^{2.1}(A,q)_{[1]} = \frac{\{A\}}{\{q\}}\left(Aq+\frac{A^2}{q^2}-\frac{A^3}{q}\right)
\ee
with $A=q^N$ and $\{x\}=x-x^{-1}$.
Note in passing, that in variance with ordinary knots, there is no symmetry under the
change $q\longrightarrow -q^{-1}$, even though the polynomial corresponds to the fundamental
representation with transposition-symmetric Young diagram.
This expression is for unreduced HOMFLY, in reduced one the first factor $[N]=\frac{\{A\}}{\{q\}}$
is omitted.
In this case there is just one Seifert cycle, thus the matrix-model fat graph contains just a single
vertex with two attached edges:

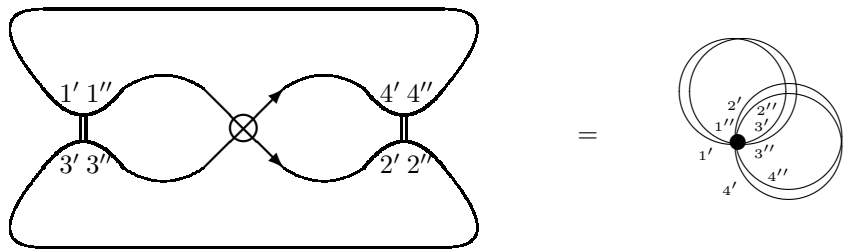
\begin{picture}(300,120)(-50,10)
  \put(100,50){
    \put(0,0){
      \thicklines
      \put(0,0){\vector(1,1){30}}
      \put(0,30){\vector(1,-1){30}}
      \put(15,15){\circle{10}}
    }
    \put(60,0){
      \thicklines
      \qbezier(0,0)(15,20)(30,0)
      \qbezier(0,30)(15,10)(30,30)
      \put(14,10){\line(0,1){10}}
      \put(16,10){\line(0,1){10}}
      \put(6,-2){\mbox{$2'$}}
      \put(16,-2){\mbox{$2''$}}
      \put(6,25){\mbox{$4'$}}
      \put(16,25){\mbox{$4''$}}
    }
    \put(-60,0){
      \thicklines

    \qbezier(0,0)(15,20)(30,0)
      \qbezier(0,30)(15,10)(30,30)
      \put(14,10){\line(0,1){10}}
      \put(16,10){\line(0,1){10}}
      \put(6,-2){\mbox{$3'$}}
      \put(16,-2){\mbox{$3''$}}
      \put(6,25){\mbox{$1'$}}
      \put(16,25){\mbox{$1''$}}
    }
    \put(30,30){\linethickness{0.2mm} \qbezier(0,0)(15,10)(30,0)}
    \put(30,0){\linethickness{0.2mm} \qbezier(0,0)(15,-10)(30,0)}
    \put(-30,30){\linethickness{0.2mm} \qbezier(0,0)(15,10)(30,0)}
    \put(-30,0){\linethickness{0.2mm} \qbezier(0,0)(15,-10)(30,0)}
    \put(-60,60){\thicklines \line(1,0){150}}
    \put(-60,-30){\thicklines \line(1,0){150}}
    \put(-60,-30){\linethickness{0.2mm} \qbezier(0,0)(-25,0)(0,30)}
    \put(-60,30){\linethickness{0.2mm} \qbezier(0,0)(-25,30)(0,30)}
    \put(90,-30){\linethickness{0.2mm} \qbezier(0,0)(25,0)(0,30)}
    \put(90,30){\linethickness{0.2mm} \qbezier(0,0)(25,30)(0,30)}
  }

  \put(240,60){\mbox{$=$}}

  \put(300,60){\circle*{6}}
  \put(302,79){\circle{40}}
  \put(298,79){\circle{40}}
  \put(319,62){\circle{40}}
  \put(319,58){\circle{40}}

{\tiny
  \put(285,53){\mbox{$  1'$}}
  \put(291,64){\mbox{$1''$}}
  \put(296,72){\mbox{$2'$}}
  \put(307,70){\mbox{$2''$}}
  \put(306,64){\mbox{$3'$}}
  \put(306,54){\mbox{$3''$}}
  \put(294,40){\mbox{$4'$}}
  \put(311,45){\mbox{$4''$}}
}
\end{picture}

\bigskip

According to the cabling rules from  \cite{AnoMcabling}
the 2-cabled knot diagram is

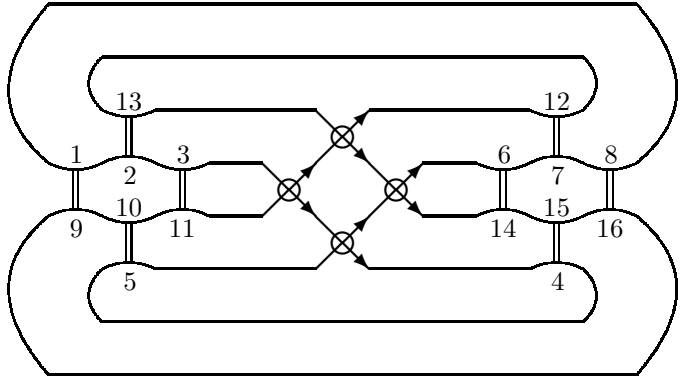
\begin{picture}(300,170)(-90,5)
  \put(20,80){
    \put(0,0){
      \linethickness{0.2mm}
      \qbezier(0,0)(10,5)(20,0)
      \put(0,20){\qbezier(0,0)(10,-5)(20,0)}
      \thinlines
      \put(9,2){\line(0,1){16}}
      \put(11,2){\line(0,1){16}}
      \put(8,20){\mbox{1}}
      \put(8,-8){\mbox{9}}
    }
    \put(40,0){
      \linethickness{0.2mm}
      \qbezier(0,0)(10,5)(20,0)
      \put(0,20){\qbezier(0,0)(10,-5)(20,0)}
      \thinlines
      \put(9,2){\line(0,1){16}}
      \put(11,2){\line(0,1){16}}
      \put(8,20){\mbox{3}}
      \put(5,-8){\mbox{11}}
    }
    \put(20,20){
      \linethickness{0.2mm}
      \qbezier(0,0)(10,5)(20,0)
      \put(0,20){\qbezier(0,0)(10,-5)(20,0)}
      \thinlines
      \put(9,2){\line(0,1){16}}
      \put(11,2){\line(0,1){16}}
      \put(5,20){\mbox{13}}
      \put(8,-8){\mbox{2}}
    }
    \put(20,-20){
      \linethickness{0.2mm}
      \qbezier(0,0)(10,5)(20,0)
      \put(0,20){\qbezier(0,0)(10,-5)(20,0)}
      \thinlines
      \put(9,2){\line(0,1){16}}
      \put(11,2){\line(0,1){16}}
      \put(5,20){\mbox{10}}
      \put(8,-8){\mbox{5}}
    }
  }
  \put(180,80){
    \put(0,0){
      \linethickness{0.2mm}
      \qbezier(0,0)(10,5)(20,0)
      \put(0,20){\qbezier(0,0)(10,-5)(20,0)}
      \thinlines
      \put(9,2){\line(0,1){16}}
      \put(11,2){\line(0,1){16}}
      \put(8,20){\mbox{6}}
      \put(5,-8){\mbox{14}}
    }
    \put(40,0){
      \linethickness{0.2mm}
      \qbezier(0,0)(10,5)(20,0)
      \put(0,20){\qbezier(0,0)(10,-5)(20,0)}
      \thinlines
      \put(9,2){\line(0,1){16}}
      \put(11,2){\line(0,1){16}}
      \put(8,20){\mbox{8}}
      \put(5,-8){\mbox{16}}
    }
    \put(20,20){
      \linethickness{0.2mm}
      \qbezier(0,0)(10,5)(20,0)
      \put(0,20){\qbezier(0,0)(10,-5)(20,0)}
      \thinlines
      \put(9,2){\line(0,1){16}}
      \put(11,2){\line(0,1){16}}
      \put(5,20){\mbox{12}}
      \put(8,-8){\mbox{7}}
    }
    \put(20,-20){
      \linethickness{0.2mm}
      \qbezier(0,0)(10,5)(20,0)
      \put(0,20){\qbezier(0,0)(10,-5)(20,0)}
      \thinlines
      \put(9,2){\line(0,1){16}}
      \put(11,2){\line(0,1){16}}
      \put(5,20){\mbox{15}}
      \put(8,-8){\mbox{4}}
    }
  }
  \put(100,80){
    \put(0,0){
      \thicklines
      \put(0,0){\vector(1,1){20}}
      \put(0,20){\vector(1,-1){20}}
      \put(10,10){\circle{8}}
    }
    \put(40,0){
      \thicklines
      \put(0,0){\vector(1,1){20}}
      \put(0,20){\vector(1,-1){20}}
      \put(10,10){\circle{8}}
    }
    \put(20,20){
      \thicklines
      \put(0,0){\vector(1,1){20}}
      \put(0,20){\vector(1,-1){20}}
      \put(10,10){\circle{8}}
    }
    \put(20,-20){
      \thicklines
      \put(0,0){\vector(1,1){20}}
      \put(0,20){\vector(1,-1){20}}
      \put(10,10){\circle{8}}
    }
    \put(-40,-20){
      \thicklines
      \put(0,0){\line(1,0){60}}
      \put(20,20){\line(1,0){20}}
      \put(20,40){\line(1,0){20}}
      \put(0,60){\line(1,0){60}}
    }
    \put(40,-20){
      \thicklines
      \put(0,0){\line(1,0){60}}
      \put(20,20){\line(1,0){20}}
      \put(20,40){\line(1,0){20}}
      \put(0,60){\line(1,0){60}}
    }
  }
  \put(20,20){
    \put(0,0){\thicklines \line(1,0){220}}
    \put(220,0){\linethickness{0.2mm} \qbezier(0,0)(30,35)(0,60)}
    \linethickness{0.2mm} \qbezier(0,0)(-30,35)(0,60)
  }
  \put(20,160){
    \put(0,0){\thicklines \line(1,0){220}}
    \put(220,0){\linethickness{0.2mm} \qbezier(0,0)(30,-35)(0,-60)}
    \linethickness{0.2mm} \qbezier(0,0)(-30,-35)(0,-60)
  }
  \put(40,40){
    \put(0,0){\thicklines \line(1,0){180}}
    \put(180,0){\linethickness{0.2mm} \qbezier(0,0)(10,10)(0,20)}
    \linethickness{0.2mm} \qbezier(0,0)(-10,10)(0,20)
  }
  \put(40,140){
    \put(0,0){\thicklines \line(1,0){180}}
    \put(180,0){\linethickness{0.2mm} \qbezier(0,0)(10,-10)(0,-20)}
    \linethickness{0.2mm} \qbezier(0,0)(-10,-10)(0,-20)
  }
\end{picture}

\noindent
This time we have two Seifert cycles and the fat graph consists of two vertices,
connected by eight edges.

Hypercube in this case is quite big. It is 8-dimensional and number of vertices in each slice
(subset of hypercube vertices with same number of edges kept) is equal, respectively,
to  1, 8, 28, 56, 70, 56, 28, 8, 1. In total there are $2^8=256$ vertices.
Describing them all explicitly is not very illustrative
and we do this only for the two simplest cases.

In the initial (Seifert) vertex of the hypercube
stands the resolved knot diagram with no edges -- we have just two Seifert cycles,
i.e. the {\it fat graph} consists of two disconnected vertices.
Each contributes a factor of $[N]$ and the answer for the dimension is $[N]^2$.

In the next slice of the hypercube just one edge is kept in the fat graph.
Since in this case every edge on the 2-cabled picture actually connects two Seifert cycles,
the fat graph is the same for all of them and the corresponding dimension is $[N][N-1]$.

Making the same analysis for all other slices of the hypercube
(with the help of computer program, of course)
we obtain the answer for the 2-cabled virtual trefoil:
\begin{align}
  H_{[1]\times[1]}^{\rm naive} = &\frac{q^{5 N-7} [N]}{(q-1) (q+1)} \Big(
   q^{6N+4} \ \     -q^{5 N+1}-2 q^{5 N+3}+q^{5 N+5}+   \\ \notag &
+q^{4 N}+2 q^{4 N+2}-2 q^{4 N+4}-q^{4N+6}-3 q^{4N+8}+q^{4N+10} \ \
 +q^{3 N+3}+3 q^{3 N+5}+q^{3N+7}-q^{3 N+9}  -   \\ \notag &
  -q^{2 N}-2 q^{2N+4}+2 q^{2 N+6}+q^{2 N+8}  +2 q^{2 N+10}   \ \    -q^{N+7}-2 q^{N+9}+q^{N+11}
   \ \  -q^{10}+q^{12}-q^{14} \Big)
\end{align}

Direct check shows, that at $N = 2$ this \textit{does not} reproduce the known answer for the 2-cabled Jones
from \cite{virtkatlas}
\begin{align}
  H^{\rm naive}_{[1]\times[1]} \Big{|}_{N=2}\ =  \ [2] q^4 \left (
  q^{15} - q^{13} - q^{11} - q^9 + 2 q^7 + q^5 + q^3 \right)
  \ \  \ \notag \neq \ \ \ [2] \left ( q^{15} - q^{13} - q^{11} - q^9 + q^7 + 2 q^5 + q \right )
\end{align}
This is, in fact, expected, because the cabling procedure that we used is too naive
and differs from the right one.
The point is that if cabling is performed by doubling the line in the planar diagram,
the two wires in the cable get linked even for the unknot -- and what we get is rather
a Hopf link or even more complicated 2-strand links $[2,2k]$ instead of two disconnected unknots.
To get the right quantity, which for ordinary knots possesses decomposition
\be
H_{[1]\times[1]} = H_{[2]} +H_{[11]}
\label{Hca2deco}
\ee
into two colored HOMFLY polynomials one needs to perform additional untwisting.
It, however, involves insertion of additional vertices and thus complicates the calculation.
For ordinary knots one could cure the problem simply by inserting additional framing factors
at the r.h.s. of (\ref{Hca2deco}) -- and this was the trick, used in \cite{AnoMcabling}.
However, for virtual knots it does not work so simple -- and a new calculation with
additional vertices is required.

\subsection{Link-free cabling for 2-cabled virtual trefoil}

What one needs to do is to introduce into the cabled knot a mutual ``twist'' of cable constituents,
such that the linking number of any two cable components vanishes.
Beginning from triple cables, this is not the whole story, because one should
also eliminate situations like Borromean rings, which are linked, despite pairwise
linking is absent. However, for double-wire cables the simple untwisting is enough.

In the case of virtual trefoil this amounts to insertion of 4 black crossings as shown on the picture

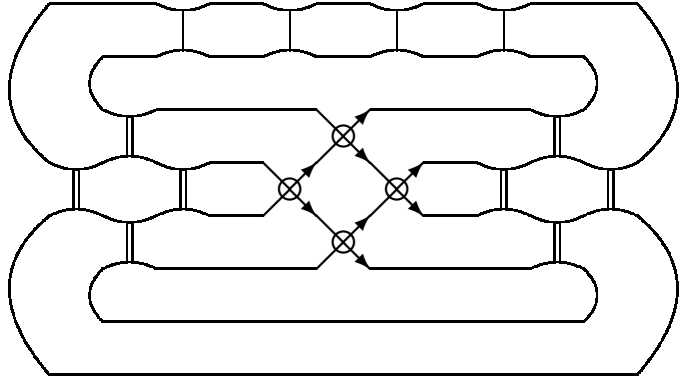
\begin{picture}(300,200)(-100,-10)
  \put(20,80){
    \put(0,0){
      \linethickness{0.2mm}
      \qbezier(0,0)(10,5)(20,0)
      \put(0,20){\qbezier(0,0)(10,-5)(20,0)}
      \thinlines
      \put(9,2){\line(0,1){16}}
      \put(11,2){\line(0,1){16}}
    }
    \put(40,0){
      \linethickness{0.2mm}
      \qbezier(0,0)(10,5)(20,0)
      \put(0,20){\qbezier(0,0)(10,-5)(20,0)}
      \thinlines
      \put(9,2){\line(0,1){16}}
      \put(11,2){\line(0,1){16}}
    }
    \put(20,20){
      \linethickness{0.2mm}
      \qbezier(0,0)(10,5)(20,0)
      \put(0,20){\qbezier(0,0)(10,-5)(20,0)}
      \thinlines
      \put(9,2){\line(0,1){16}}
      \put(11,2){\line(0,1){16}}
    }
    \put(20,-20){
      \linethickness{0.2mm}
      \qbezier(0,0)(10,5)(20,0)
      \put(0,20){\qbezier(0,0)(10,-5)(20,0)}
      \thinlines
      \put(9,2){\line(0,1){16}}
      \put(11,2){\line(0,1){16}}
    }
  }
  \put(180,80){
    \put(0,0){
      \linethickness{0.2mm}
      \qbezier(0,0)(10,5)(20,0)
      \put(0,20){\qbezier(0,0)(10,-5)(20,0)}
      \thinlines
      \put(9,2){\line(0,1){16}}
      \put(11,2){\line(0,1){16}}
    }
    \put(40,0){
      \linethickness{0.2mm}
      \qbezier(0,0)(10,5)(20,0)
      \put(0,20){\qbezier(0,0)(10,-5)(20,0)}
      \thinlines
      \put(9,2){\line(0,1){16}}
      \put(11,2){\line(0,1){16}}
    }
    \put(20,20){
      \linethickness{0.2mm}
      \qbezier(0,0)(10,5)(20,0)
      \put(0,20){\qbezier(0,0)(10,-5)(20,0)}
      \thinlines
      \put(9,2){\line(0,1){16}}
      \put(11,2){\line(0,1){16}}
    }
    \put(20,-20){
      \linethickness{0.2mm}
      \qbezier(0,0)(10,5)(20,0)
      \put(0,20){\qbezier(0,0)(10,-5)(20,0)}
      \thinlines
      \put(9,2){\line(0,1){16}}
      \put(11,2){\line(0,1){16}}
    }
  }
  \put(100,80){
    \put(0,0){
      \thicklines
      \put(0,0){\vector(1,1){20}}
      \put(0,20){\vector(1,-1){20}}
      \put(10,10){\circle{8}}
    }
    \put(40,0){
      \thicklines
      \put(0,0){\vector(1,1){20}}
      \put(0,20){\vector(1,-1){20}}
      \put(10,10){\circle{8}}
    }
    \put(20,20){
      \thicklines
      \put(0,0){\vector(1,1){20}}
      \put(0,20){\vector(1,-1){20}}
      \put(10,10){\circle{8}}
    }
    \put(20,-20){
      \thicklines
      \put(0,0){\vector(1,1){20}}
      \put(0,20){\vector(1,-1){20}}
      \put(10,10){\circle{8}}
    }
    \put(-40,-20){
      \thicklines
      \put(0,0){\line(1,0){60}}
      \put(20,20){\line(1,0){20}}
      \put(20,40){\line(1,0){20}}
      \put(0,60){\line(1,0){60}}
    }
    \put(40,-20){
      \thicklines
      \put(0,0){\line(1,0){60}}
      \put(20,20){\line(1,0){20}}
      \put(20,40){\line(1,0){20}}
      \put(0,60){\line(1,0){60}}
    }
  }
  \put(20,20){
    \put(0,0){\thicklines \line(1,0){220}}
    \put(220,0){\linethickness{0.2mm} \qbezier(0,0)(30,35)(0,60)}
    \linethickness{0.2mm} \qbezier(0,0)(-30,35)(0,60)
  }
  \put(20,160){
    \put(0,0){
      \thicklines
      \put(0,0){\line(1,0){40}}
      \put(60,0){\line(1,0){20}}
      \put(100,0){\line(1,0){20}}
      \put(140,0){\line(1,0){20}}
      \put(180,0){\line(1,0){40}}
    }
    \put(220,0){\linethickness{0.2mm} \qbezier(0,0)(30,-35)(0,-60)}
    \linethickness{0.2mm} \qbezier(0,0)(-30,-35)(0,-60)
  }
  \put(40,40){
    \put(0,0){\thicklines \line(1,0){180}}
    \put(180,0){\linethickness{0.2mm} \qbezier(0,0)(10,10)(0,20)}
    \linethickness{0.2mm} \qbezier(0,0)(-10,10)(0,20)
  }
  \put(40,140){
    \put(0,0){
      \thicklines
      \put(0,0){\line(1,0){20}}
      \put(40,0){\line(1,0){20}}
      \put(80,0){\line(1,0){20}}
      \put(120,0){\line(1,0){20}}
      \put(160,0){\line(1,0){20}}
    }
    \put(180,0){\linethickness{0.2mm} \qbezier(0,0)(10,-10)(0,-20)}
    \linethickness{0.2mm} \qbezier(0,0)(-10,-10)(0,-20)
  }
  \put(60,140){
    \linethickness{0.2mm}
    \qbezier(0,0)(10,5)(20,0)
    \put(0,20){\qbezier(0,0)(10,-5)(20,0)}
    \thinlines
    \put(10,2){\line(0,1){16}}
  }
  \put(100,140){
    \linethickness{0.2mm}
    \qbezier(0,0)(10,5)(20,0)
    \put(0,20){\qbezier(0,0)(10,-5)(20,0)}
    \thinlines
    \put(10,2){\line(0,1){16}}
  }
  \put(140,140){
    \linethickness{0.2mm}
    \qbezier(0,0)(10,5)(20,0)
    \put(0,20){\qbezier(0,0)(10,-5)(20,0)}
    \thinlines
    \put(10,2){\line(0,1){16}}
  }
  \put(180,140){
    \linethickness{0.2mm}
    \qbezier(0,0)(10,5)(20,0)
    \put(0,20){\qbezier(0,0)(10,-5)(20,0)}
    \thinlines
    \put(10,2){\line(0,1){16}}
  }
\end{picture}

Then the answer is
\begin{align}
  \frac{H_{[1]\times[1]}^{2.1}}{[N]} =
  \frac{q^{N-8}}{(q-1) (q+1)} \Big( & q^{6 N+1}-q^{6 N+3}+q^{6 N+5}-q^{6 N+7}+q^{6 N+9}
  \\ \notag
  & -q^{5 N} -q^{5 N+6}-q^{5 N+8}+q^{5 N+10}
  \\ \notag
  & +q^{4 N+1}-2 q^{4 N+3}+2 q^{4 N+5}-q^{4 N+9}-q^{4 N+11}-2 q^{4 N+13}+q^{4 N+15}
  \\ \notag
  & +q^{3 N+2}+q^{3 N+4}+q^{3 N+8}+2 q^{3 N+10}-q^{3 N+14}
  \\ \notag
  & -q^{2 N+3}-q^{2 N+9}+3 q^{2 N+11} +2 q^{2 N+15}-q^{2 N+17}
  \\ \notag
  & -q^{N+6}-q^{N+12}-q^{N+14}+q^{N+16}
  & \\ \notag & -q^{13}\Big)
\end{align}
and this time it {\it does} reproduce the cabled Jones from \cite{virtkatlas}
\begin{align}
  H_{[1]\times[1]}^{2.1} \Big{|}_{N=2} = & [2] \left ( q^{15} - q^{13} - q^{11} - q^9 + q^7 + 2 q^5 + q \right )
\end{align}
which possesses spectacular decomposition properties {\it a la} (\ref{Hca2deco}),
see sec.7 of the second paper in \cite{DM3virt}, implying the possibility to define
colored polynomials for virtual knots(!).

\subsection{2-cabled virtual knot 3.2}

Let's consider one more example. This is a 3.2 knot in the terminology of \cite{virtkatlas}

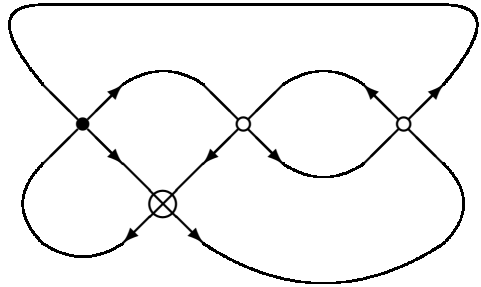
\begin{picture}(300,120)(-100,0)
  \put(100,50){
    \put(-30,-30){
      \thicklines
      \put(30,30){\vector(-1,-1){30}}
      \put(0,30){\vector(1,-1){30}}
      \put(15,15){\circle{10}}
    }
    \put(0,0){
      \thicklines
      \put(13,13){\vector(-1,-1){13}}
      \put(17,17){\line(1,1){13}}
      \put(0,30){\line(1,-1){13}}
      \put(17,13){\vector(1,-1){13}}
      \put(15,15){\circle{5}}
    }
    \put(60,0){
      \thicklines
      \put(17,17){\vector(1,1){13}}
      \put(0,0){\line(1,1){13}}
      \put(13,17){\vector(-1,1){13}}
      \put(17,13){\line(1,-1){13}}
      \put(15,15){\circle{5}}
    }
    \put(-60,0){
      \thicklines
      \put(0,0){\line(1,1){13}}
      \put(17,17){\vector(1,1){13}}
      \put(0,30){\line(1,-1){13}}
      \put(17,13){\vector(1,-1){13}}
      \put(15,15){\circle*{5}}
    }
    \put(30,30){\linethickness{0.2mm} \qbezier(0,0)(15,10)(30,0)}
    \put(30,0){\linethickness{0.2mm} \qbezier(0,0)(15,-10)(30,0)}
    \put(-30,30){\linethickness{0.2mm} \qbezier(0,0)(15,10)(30,0)}
    \put(-60,60){\thicklines \line(1,0){150}}
    \linethickness{0.2mm}
    \put(-60,-30){\qbezier(0,0)(15,-10)(30,0)}
    \put(0,-30){\qbezier(0,0)(45,-30)(90,0)}
    \put(-60,-30){\linethickness{0.2mm} \qbezier(0,0)(-15,15)(0,30)}
    \put(-60,30){\linethickness{0.2mm} \qbezier(0,0)(-25,30)(0,30)}
    \put(90,-30){\linethickness{0.2mm} \qbezier(0,0)(15,15)(0,30)}
    \put(90,30){\linethickness{0.2mm} \qbezier(0,0)(25,30)(0,30)}
  }
\end{picture}

\noindent
We see, that there are two white vertices and one black, thus to unlink the 2-cable
we must additionally insert two black intersections.

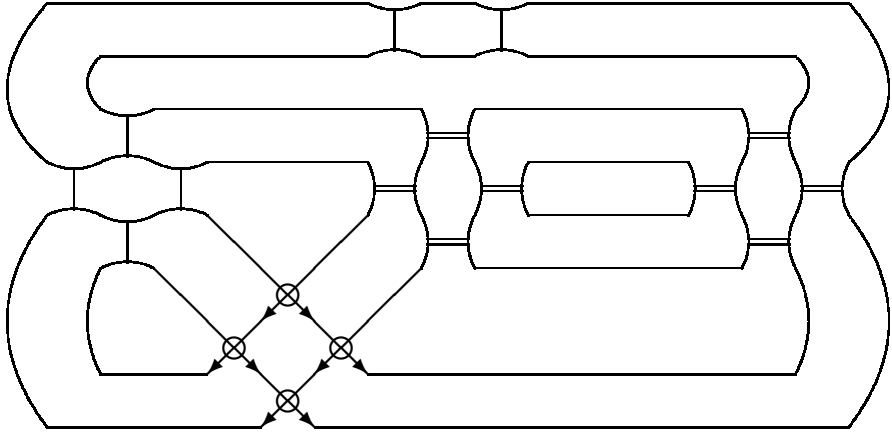
\begin{picture}(300,200)(-70,-20)
  \put(20,80){
    \put(0,0){
      \linethickness{0.2mm}
      \qbezier(0,0)(10,5)(20,0)
      \put(0,20){\qbezier(0,0)(10,-5)(20,0)}
      \thinlines
      \put(10,2){\line(0,1){16}}
    }
    \put(40,0){
      \linethickness{0.2mm}
      \qbezier(0,0)(10,5)(20,0)
      \put(0,20){\qbezier(0,0)(10,-5)(20,0)}
      \thinlines
      \put(10,2){\line(0,1){16}}
    }
    \put(20,20){
      \linethickness{0.2mm}
      \qbezier(0,0)(10,5)(20,0)
      \put(0,20){\qbezier(0,0)(10,-5)(20,0)}
      \thinlines
      \put(10,2){\line(0,1){16}}
    }
    \put(20,-20){
      \linethickness{0.2mm}
      \qbezier(0,0)(10,5)(20,0)
      \put(0,20){\qbezier(0,0)(10,-5)(20,0)}
      \thinlines
      \put(10,2){\line(0,1){16}}
    }
  }
  \put(140,80){
    \put(0,0){
      \linethickness{0.2mm}
      \qbezier(0,0)(5,10)(0,20)
      \put(20,0){\qbezier(0,0)(-5,10)(0,20)}
      \thinlines
      \put(2,9){\line(1,0){16}}
      \put(2,11){\line(1,0){16}}
    }
    \put(40,0){
      \linethickness{0.2mm}
      \qbezier(0,0)(5,10)(0,20)
      \put(20,0){\qbezier(0,0)(-5,10)(0,20)}
      \thinlines
      \put(2,9){\line(1,0){16}}
      \put(2,11){\line(1,0){16}}
    }
    \put(20,20){
      \linethickness{0.2mm}
      \qbezier(0,0)(5,10)(0,20)
      \put(20,0){\qbezier(0,0)(-5,10)(0,20)}
      \thinlines
      \put(2,9){\line(1,0){16}}
      \put(2,11){\line(1,0){16}}
    }
    \put(20,-20){
      \linethickness{0.2mm}
      \qbezier(0,0)(5,10)(0,20)
      \put(20,0){\qbezier(0,0)(-5,10)(0,20)}
      \thinlines
      \put(2,9){\line(1,0){16}}
      \put(2,11){\line(1,0){16}}
    }
  }
  \put(260,80){
    \put(0,0){
      \linethickness{0.2mm}
      \qbezier(0,0)(5,10)(0,20)
      \put(20,0){\qbezier(0,0)(-5,10)(0,20)}
      \thinlines
      \put(2,9){\line(1,0){16}}
      \put(2,11){\line(1,0){16}}
    }
    \put(40,0){
      \linethickness{0.2mm}
      \qbezier(0,0)(5,10)(0,20)
      \put(20,0){\qbezier(0,0)(-5,10)(0,20)}
      \thinlines
      \put(2,9){\line(1,0){16}}
      \put(2,11){\line(1,0){16}}
    }
    \put(20,20){
      \linethickness{0.2mm}
      \qbezier(0,0)(5,10)(0,20)
      \put(20,0){\qbezier(0,0)(-5,10)(0,20)}
      \thinlines
      \put(2,9){\line(1,0){16}}
      \put(2,11){\line(1,0){16}}
    }
    \put(20,-20){
      \linethickness{0.2mm}
      \qbezier(0,0)(5,10)(0,20)
      \put(20,0){\qbezier(0,0)(-5,10)(0,20)}
      \thinlines
      \put(2,9){\line(1,0){16}}
      \put(2,11){\line(1,0){16}}
    }
  }
  \put(80,20){
    \put(0,0){
      \thicklines
      \put(20,20){\vector(-1,-1){20}}
      \put(0,20){\vector(1,-1){20}}
      \put(10,10){\circle{8}}
    }
    \put(40,0){
      \thicklines
      \put(20,20){\vector(-1,-1){20}}
      \put(0,20){\vector(1,-1){20}}
      \put(10,10){\circle{8}}
    }
    \put(20,20){
      \thicklines
      \put(20,20){\vector(-1,-1){20}}
      \put(0,20){\vector(1,-1){20}}
      \put(10,10){\circle{8}}
    }
    \put(20,-20){
      \thicklines
      \put(20,20){\vector(-1,-1){20}}
      \put(0,20){\vector(1,-1){20}}
      \put(10,10){\circle{8}}
    }
  }

  \put(20,160){
    \put(0,0){
      \thicklines
      \put(0,0){\line(1,0){120}}
      \put(140,0){\line(1,0){20}}
      \put(180,0){\line(1,0){120}}
    }
    \put(300,0){\linethickness{0.2mm} \qbezier(0,0)(30,-35)(0,-60)}
    \linethickness{0.2mm} \qbezier(0,0)(-30,-35)(0,-60)
  }
  \put(40,140){
    \put(0,0){
      \thicklines
      \put(0,0){\line(1,0){100}}
      \put(120,0){\line(1,0){20}}
      \put(160,0){\line(1,0){100}}
    }
    \put(260,0){\linethickness{0.2mm} \qbezier(0,0)(10,-10)(0,-20)}
    \linethickness{0.2mm} \qbezier(0,0)(-10,-10)(0,-20)
  }

  \put(20,80){
    \put(40,40){\line(1,0){100}}
    \put(60,20){\line(1,0){60}}
  }
  \put(140,80){
    \put(40,40){\line(1,0){100}}
    \put(60,20){\line(1,0){60}}
  }
  \put(140,20){
    \put(40,40){\line(1,0){100}}
    \put(60,60){\line(1,0){60}}
  }

  \put(140,140){
    \linethickness{0.2mm}
    \qbezier(0,0)(10,5)(20,0)
    \put(0,20){\qbezier(0,0)(10,-5)(20,0)}
    \thinlines
    \put(10,2){\line(0,1){16}}
  }
  \put(180,140){
    \linethickness{0.2mm}
    \qbezier(0,0)(10,5)(20,0)
    \put(0,20){\qbezier(0,0)(10,-5)(20,0)}
    \thinlines
    \put(10,2){\line(0,1){16}}
  }

  \thicklines
  \put(80,80){\line(1,-1){20}}
  \put(60,60){\line(1,-1){20}}
  \put(120,60){\line(1,1){20}}
  \put(140,40){\line(1,1){20}}

  \put(140,20){\line(1,0){160}}
  \put(120,0){\line(1,0){200}}
  \put(20,0){
    \put(300,0){\linethickness{0.2mm} \qbezier(0,0)(30,40)(0,80)}
    \put(280,0){\linethickness{0.2mm} \qbezier(0,20)(10,40)(0,60)}
  }

  \put(140,20){\line(1,0){160}}
  \put(120,0){\line(1,0){200}}
  \put(20,0){
    \put(0,0){\linethickness{0.2mm} \qbezier(0,0)(-30,40)(0,80)}
    \put(20,20){\linethickness{0.2mm} \qbezier(0,0)(-10,20)(0,40)}
    \put(20,20){\line(1,0){40}}
    \put(0,0){\line(1,0){80}}
  }

\end{picture}

Then applying the ribbon-graph machinery we get
\begin{align}
H^{3.2}_{[1]\otimes[1]} =
\frac{q^{-3 N-5}}{(q-1) (q+1)} \Big(
& q^{8 N+2}-q^{8 N+4}+q^{8 N+6}-q^{8 N+8} +q^{8 N+10}
\\ \notag & -q^{7 N+1}-q^{7 N+5}
\\ \notag & +q^{6 N+2} +q^{6 N+4}-q^{6 N+10}-q^{6 N+12}
\\ \notag & -q^{5 N+1}+2 q^{5 N+5}+2 q^{5 N+7}+q^{5 N+9}
\\ \notag & + q^{4 N} -q^{4 N+2}-q^{4 N+4} -5 q^{4 N+6}+q^{4 N+10}+2 q^{4 N+12}
\\ \notag & +q^{3 N+3}+2 q^{3 N+5}-q^{3 N+7}-2 q^{3 N+9}-2 q^{3 N+11}
\\ \notag & -q^{2 N+2}+q^{2 N+4}+3 q^{2 N+8}+2 q^{2 N+10} -2 q^{2 N+12}
\\ \notag & -q^{N+7}-q^{N+9}+q^{N+11}+q^{N+13}
\\ \notag & -q^8+q^{10}-q^{12}\Big)
\end{align}

\noindent
which again reproduces at $N=2$ the cabled Jones from \cite{virtkatlas}
\begin{align}
  H_{[1]\times[1]}^{3.2} \Big{|}_{N=2} = & [2] \left ( q^{13}-q^{11}-q^9+2 q^5+q^3-q+q^{-5} \right )
\end{align}

\bigskip

Further details and results on cabled HOMFLY will be provided elsewhere.
In the remaining section we suggest one more application of the matrix-model/fat-graph
formalism -- which is not so technical and can be of more general value.

\section{Relation to Turaev genus}

Important characteristic of the fat graph is its topology.
In fact, the genus of our graphs appears closely related to {\it Turaev genus} \cite{Tug},
which is the upper bound of Khovanov's width \cite{upbou}.
More precisely, this relation directly concerns only link diagrams {\bf with all black vertices}.
In such case, if our fat graph is {\it cross}-planar, then the corresponding knot is necessarily {\it thin},
and for {\it thick} knots the fat graphs are obligatory {\it cross}-non-planar.
However, {\it cross}-non-planar graphs can still be associated with thin knots.

Turaev genus of the link diagram ${\cal L}$ with $n$  vertices is
\be
g_T^{\cal L} = \frac{2+n-s_{||}-s_{=}}{2}
\label{TugL}
\ee
where $s_{||}$ is the number of Seifert cycles, associated with the resolution
$\begin{picture}(60,10)(-5,2) \put(0,0){\vector(1,1){10}}\put(9,0){\vector(-1,1){10}}
\put(4.5,4.5){\circle*{4}} \put(20,4.5){\vector(1,0){15}} \put(45,0){\vector(0,1){10}}
\put(50,0){\vector(0,1){10}}\end{picture}$,
$\begin{picture}(70,10)(-5,2) \put(0,0){\vector(1,1){10}}\put(9,0){\vector(-1,1){10}}
\put(4.5,4.5){\circle{4}}  \put(20,4.5){\vector(1,0){15}}
\qbezier(45,0)(52,7)(59,0) \qbezier(45,10)(52,3)(59,10)
\put(47,8){\vector(-1,1){2}} \put(57,8){\vector(1,1){2}}
\put(45,0){\vector(1,1){2}} \put(59,0){\vector(-1,1){2}} \end{picture}$,
while $s_{=}$ is the number of cycles,  associated with the resolution
$\begin{picture}(70,10)(-5,2) \put(0,0){\vector(1,1){10}}\put(9,0){\vector(-1,1){10}}
\put(4.5,4.5){\circle*{4}}  \put(20,4.5){\vector(1,0){15}}
\qbezier(45,0)(52,7)(59,0) \qbezier(45,10)(52,3)(59,10)
\put(47,8){\vector(-1,1){2}} \put(57,8){\vector(1,1){2}}
\put(45,0){\vector(1,1){2}} \put(59,0){\vector(-1,1){2}} \end{picture}$,
$\begin{picture}(60,10)(-5,2) \put(0,0){\vector(1,1){10}}\put(9,0){\vector(-1,1){10}}
\put(4.5,4.5){\circle{4}} \put(20,4.5){\vector(1,0){15}} \put(45,0){\vector(0,1){10}}
\put(50,0){\vector(0,1){10}}\end{picture}$.
Then Turaev genus of the knot is the minimum of these quantities over all possible
diagrams, representing the knot.
Turaev genus vanishes if and only if the knot is alternating
(i.e. such that when we walk along the planar diagram of this knot and keep track of the type of intersections
we encounter, we notice, that the type alternates -- say, if at first intersection we pass above the other strand,
then at the second we necessarily pass below, at the third again above and so on).

We defined our fat graphs only for diagrams ${\cal L}$ with all black vertices.
Then $s_{||}$ and $n$ are respectively the numbers of vertices and edges of the fat graph.
If $s_{=}$ was the number of loops (boundary components)
of the ``thickening'' of the fat graph (i.e. the result of putting graph without self-intersections
on a Riemann surface, respecting orientations at vertices, and then cutting a piece out of this surface
according to the graph, such that its vertices become small disks and edges become small stripes)
, then Turaev genus (\ref{TugL}) would be exactly
the genus of the fat graph $\Gamma^{\cal L}$ (defined as genus of this surface we can put the graph on).

Of course, this can not be true for the ribbon graph, which we associate with the knot
in this paper -- because for knots the number of loops of \textit{its} ``thickening'' is always one.
However, if we first consider the ``thickening'' and then additionally twist all the stripes
(such that twisted become untwisted and vice versa), i.e. consider it as
an ordinary matrix model graph (this is what we did in discussion of cross-planarity),
the number of boundary compontents (loops)
 of this cross-``thickening'' is different and exactly coincides with $s_{=}$.
We first give general explanation of this and then further illustrate it with
a number of examples.

Transition from usual fat graph ``thickening'' to the twisted amounts
to the following change of rules for gluing stripes to disks
(arrows on short ends of stripes have to match orientation on disks boundaries)

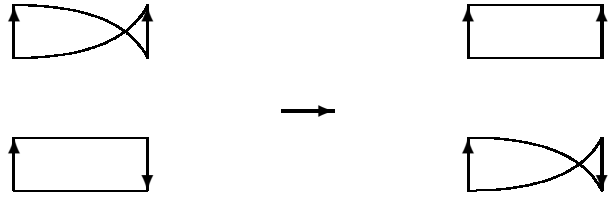
\begin{picture}(300,100)(-100,-10)
  \thicklines
  \put(0,50){
    \put(0,0){\vector(0,1){20}}
        {\linethickness{0.15mm} \qbezier(0,20)(40,20)(50,0) \qbezier(0,0)(40,0)(50,20) }
        \put(50,0){\vector(0,1){20}}
  }
  \put(0,0){
    \thicklines
    \put(0,0){\vector(0,1){20}}
        {\thinlines \put(0,20){\line(1,0){50}} \put(0,0){\line(1,0){50}}}
        \put(50,20){\vector(0,-1){20}}
  }
  \put(100,30){\vector(1,0){20}}
  \put(170,0) {
    \thicklines
    \put(0,50){
      \put(0,0){\vector(0,1){20}}
          {\thinlines \put(0,20){\line(1,0){50}} \put(0,0){\line(1,0){50}}}
          \put(50,0){\vector(0,1){20}}
    }
    \put(0,0){
      \thicklines
      \put(0,0){\vector(0,1){20}}
          {\linethickness{0.15mm} \qbezier(0,20)(40,20)(50,0) \qbezier(0,0)(40,0)(50,20) }
          \put(50,20){\vector(0,-1){20}}
    }
  }
\end{picture}

\noindent
This change clearly preserves number of edges and vertices of the graphs. However,
edges now look exactly like resolutions
$\begin{picture}(20,10)(45,2)
\qbezier(45,0)(52,7)(59,0) \qbezier(45,10)(52,3)(59,10)
\put(47,8){\vector(-1,1){2}} \put(57,8){\vector(1,1){2}}
\put(45,0){\vector(1,1){2}} \put(59,0){\vector(-1,1){2}} \end{picture}$
Hence  the number of loops (boundaries) of this new ``thickening''
is precisely the same as $s_{=}$.

\bigskip

Also, $s_{=}$ is directly related to the classical ($q=1$) dimension at $N=2$:
\be
\text{dim}_\Gamma(N=2,q=1) = 2^{s_{=}(\Gamma)}
\ee
This is simply because at $N=2$ and $q=1$ \ \
$\begin{picture}(70,10)(-5,2)
\put(0,0){\line(0,1){10}} \put(7,0){\line(0,1){10}}
\put(11,3){\mbox{$-$}}
\put(20,0){\line(1,1){10}}\put(29,0){\line(-1,1){10}}
\put(34,3){\mbox{$=$}}
\qbezier(45,0)(52,7)(59,0) \qbezier(45,10)(52,3)(59,10)
\end{picture}$,
what is the key to truncation from the general formalism of \cite{DM3}
to original Kauffman-Khovanov calculus at $N=2$.
Therefore what we will test in our examples is the relation
\be
\text{dim}_\Gamma(N=2,q=1) = 2^{\text{loops of} \ \Gamma}
\label{dimloops}
\ee

\bigskip

{\bf Ex.1: \ generalized trees (cross-planar)} \ \ \
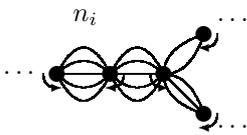
\begin{picture}(80,20)(180,55)
  \put(-20,0){\mbox{$\ldots$}}
  \put(0,0){
    \put(0,0){\circle*{6}}
    \qbezier(-5,0)(-5,-5)(0,-5) \put(2,-5){\vector(1,0){0}}
  }
  \put(20,0){
    \put(0,0){\circle*{6}}
    \put(0,0){\qbezier(5,0)(5,-5)(0,-5) \put(-2,-5){\vector(-1,0){0}}}
  }
  \put(40,0){
    \put(0,0){\circle*{6}}
    \qbezier(-5,0)(-5,-5)(0,-5) \put(2,-5){\vector(1,0){0}}
  }
\qbezier(0,0)(10,20)(20,0)\qbezier(0,0)(10,10)(20,0)\qbezier(0,0)(10,0)(20,0)
\qbezier(0,0)(10,-10)(20,0)\qbezier(0,0)(10,-20)(20,0)
\qbezier(40,0)(30,20)(20,0)\qbezier(40,0)(30,10)(20,0)\qbezier(40,0)(30,0)(20,0)
\qbezier(40,0)(30,-10)(20,0)\qbezier(40,0)(30,-20)(20,0)
\put(55,15){
  \put(0,0){\circle*{6}}
  \put(0,0){\qbezier(5,0)(5,-5)(0,-5) \put(-2,-5){\vector(-1,0){0}}}
}
\put(60,20){\mbox{$\ldots$}}
\put(55,-15){
  \put(0,0){\circle*{6}}
  \put(0,0){\qbezier(5,0)(5,-5)(0,-5) \put(-2,-5){\vector(-1,0){0}}}
}
\put(60,-20){\mbox{$\ldots$}}
\qbezier(40,0)(40,10)(55,15) \qbezier(40,0)(50,0)(55,15)
\qbezier(40,0)(40,-10)(55,-15) \qbezier(40,0)(50,0)(55,-15) \qbezier(40,0)(50,-10)(55,-15)
\put(6,20){\mbox{$n_i$}}
\end{picture}
\be
\# \ \text{of vertices} = s_{||} \nn \\
\# \ \text{of edges} = n = \sum_{i=1}^{s_{||}-1} n_i \nn \\
\# \ \text{of loops}= 1 + \sum_{i=1}^{s_{||}-1} (n_i-1) = n-s_{||}+2 \nn \\ \nn \\
\text{dim}
 = [N][N-1]^{s_{||}-1}\prod_{i=1}^{s_{||}-1} [2]^{n_i-1} \
\stackrel{N=2,q=1}{\longrightarrow} \ 2^{1+\sum\limits_{i=1}^{s_{||}-1}(n_i-1)} = 2^{\#\ \text{of loops}}
\ee

\bigskip

{\bf Ex.2: \ generalized cycles (cross-planar)} \ \ \

For simplicity we assume cycle has even number of vertices.

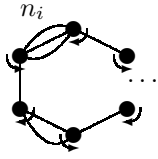
\begin{picture}(50,20)(-35,35)
  \put(40,-20){\mbox{$\ldots$}}
  \put(0,5){\mbox{$n_i$}}
  \put(0,-10){
    \put(0,0){\circle*{6}}
    \qbezier(-5,0)(-5,-5)(0,-5) \put(2,-5){\vector(1,0){0}}
  }
  \put(20,0){
    \put(0,0){\circle*{6}}
    \put(0,0){\qbezier(5,0)(5,-5)(0,-5) \put(-2,-5){\vector(-1,0){0}}}
  }
  \put(40,-10){
    \put(0,0){\circle*{6}}
    \qbezier(-5,0)(-5,-5)(0,-5) \put(2,-5){\vector(1,0){0}}
  }
  \put(0,-30){
    \put(0,0){\circle*{6}}
    \put(0,0){\qbezier(5,0)(5,-5)(0,-5) \put(-2,-5){\vector(-1,0){0}}}
  }
  \put(20,-40){
    \put(0,0){\circle*{6}}
    \qbezier(-5,0)(-5,-5)(0,-5) \put(2,-5){\vector(1,0){0}}
  }
  \put(40,-30){
    \put(0,0){\circle*{6}}
    \put(0,0){\qbezier(5,0)(5,-5)(0,-5) \put(-2,-5){\vector(-1,0){0}}}
  }
  \qbezier(0,-10)(10,-5)(20,0) \qbezier(0,-10)(10,5)(20,0) \qbezier(0,-10)(10,-15)(20,0)
  \qbezier(20,0)(30,-5)(40,-10)
  \qbezier(0,-10)(0,-20)(0,-30)
  \qbezier(0,-30)(10,-35)(20,-40) \qbezier(0,-30)(10,-50)(20,-40)
  \qbezier(20,-40)(30,-35)(40,-30)
\end{picture}
\be
\# \ \text{of vertices} = s_{||} \nn \\
\# \ \text{of edges} = n = \sum_{i=1}^{s_{||}-1} n_i \nn \\
\# \ \text{of loops}= 2 + \sum_{i=1}^{s_{||}-1} (n_i-1) = n-s_{||}+3 \nn \\ \nn \\
\text{dim}
= \left([N]^2 + [N-2] [N] \sum_{k=0}^{(s_{||}-2)/2} [N-1]^{2k} \right)\prod_{i=1}^{s_{||}-1} [2]^{n_i-1} \
\stackrel{N=2,q=1}{\longrightarrow} \ 2^{2+\sum\limits_{i=1}^{s_{||}-1}(n_i-1)} = 2^{\#\ \text{of loops}}
\ee

\bigskip

{\bf Ex.3:\ the first non-cross-planar diagram} \ \ \
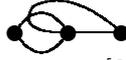
\begin{picture}(50,20)(-5,-5)
\put(0,0){\circle*{6}} \put(20,0){\circle*{6}} \put(40,0){\circle*{6}}
\qbezier(0,0)(10,15)(20,0)\qbezier(0,0)(10,-15)(20,0)
\qbezier(20,0)(30,0)(40,0)\qbezier(20,0)(0,0)(7,10)\qbezier(7,10)(21,17)(40,0)
\end{picture}

Here $n=4$, $s_{||}=3$, the number of loops is $1$,
 dimension is $[N][N-1]^2 + [N-2][2][N][N-1] \longrightarrow 2 = 2^{\#\ \text{of loops}}$.
This diagram is associated with the knot diagram, which describes trefoil $3_1$
(as a 3-strand torus knot)
and the figure-eight knot $4_1$, which are both thin.
For the trefoil there is also another possible diagram -- the tree
$\begin{picture}(30,20)(-5,-2)
  \put(0,0){
    \put(0,0){\circle*{6}}
    \put(0,0){\qbezier(-5,0)(-5,-5)(0,-5) \put(2,-5){\vector(1,0){0}}}
  }\put(20,0){
    \put(0,0){\circle*{6}}
    \put(0,0){\qbezier(5,0)(5,-5)(0,-5) \put(-2,-5){\vector(-1,0){0}}}
  }\qbezier(0,0)(10,10)(20,0)\qbezier(0,0)(10,0)(20,0)
\qbezier(0,0)(10,-10)(20,0)\end{picture}$,
representing it as a 2-strand torus knot, which is \textit{cross}-planar --
and Turaev genus of the trefoil picks the minimal value, which is zero.
For figure-eight, however, only two of the four vertices in the link
diagram are black, and when white vertices are present, the formulas for $s_{||}$
and $s_{=}$ are different -- it is easy to check that for such coloring Turaev index
of the diagram is actually zero.

Topology of this fat graph does not change if we eliminate one or two of its non-central
vertices. However, this is not so innocent at the level of knot diagram -- which
now represents virtual knots and classical dimensions can get negative, even at $N=2$.

\bigskip

{\bf Ex 4. The first thick knot $8_{19}$:} This is a three-strand torus knot,
thus all the vertices in the knot diagram are black, the fat graph has $s_{||}=3$ vertices
and eight edges (braid, as usual, is assumed to be closed from right to left)

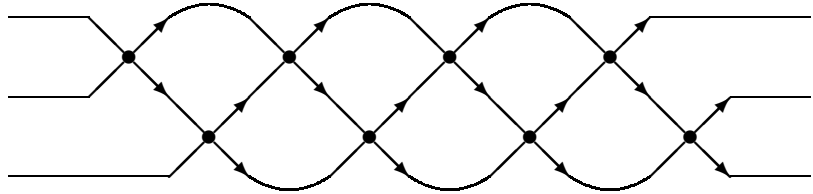
\begin{picture}(300,90)(-100,0)
  \put(100,50){
    \put(-60,0){
      \thicklines
      \put(0,0){\line(1,1){13}}
      \put(17,17){\vector(1,1){13}}
      \put(0,30){\line(1,-1){13}}
      \put(17,13){\vector(1,-1){13}}
      \put(15,15){\circle*{5}}
    }
    \put(-30,-30){
      \thicklines
      \put(0,0){\line(1,1){13}}
      \put(17,17){\vector(1,1){13}}
      \put(0,30){\line(1,-1){13}}
      \put(17,13){\vector(1,-1){13}}
      \put(15,15){\circle*{5}}
    }
    \put(60,0){
      \put(-60,0){
        \thicklines
        \put(0,0){\line(1,1){13}}
        \put(17,17){\vector(1,1){13}}
        \put(0,30){\line(1,-1){13}}
        \put(17,13){\vector(1,-1){13}}
        \put(15,15){\circle*{5}}
      }
      \put(-30,-30){
        \thicklines
        \put(0,0){\line(1,1){13}}
        \put(17,17){\vector(1,1){13}}
        \put(0,30){\line(1,-1){13}}
        \put(17,13){\vector(1,-1){13}}
        \put(15,15){\circle*{5}}
      }
    }
    \put(120,0) {
      \put(-60,0){
        \thicklines
        \put(0,0){\line(1,1){13}}
        \put(17,17){\vector(1,1){13}}
        \put(0,30){\line(1,-1){13}}
        \put(17,13){\vector(1,-1){13}}
        \put(15,15){\circle*{5}}
      }
      \put(-30,-30){
        \thicklines
        \put(0,0){\line(1,1){13}}
        \put(17,17){\vector(1,1){13}}
        \put(0,30){\line(1,-1){13}}
        \put(17,13){\vector(1,-1){13}}
        \put(15,15){\circle*{5}}
      }
      \put(60,0){
        \put(-60,0){
          \thicklines
          \put(0,0){\line(1,1){13}}
          \put(17,17){\vector(1,1){13}}
          \put(0,30){\line(1,-1){13}}
          \put(17,13){\vector(1,-1){13}}
          \put(15,15){\circle*{5}}
        }
        \put(-30,-30){
          \thicklines
          \put(0,0){\line(1,1){13}}
          \put(17,17){\vector(1,1){13}}
          \put(0,30){\line(1,-1){13}}
          \put(17,13){\vector(1,-1){13}}
          \put(15,15){\circle*{5}}
        }
      }
    }
    \put(30,30){\linethickness{0.2mm} \qbezier(0,0)(15,10)(30,0)}
    \put(-30,30){\linethickness{0.2mm} \qbezier(0,0)(15,10)(30,0)}
    \put(90,30){\linethickness{0.2mm} \qbezier(0,0)(15,10)(30,0)}
    \put(60,-30){\linethickness{0.2mm} \qbezier(0,0)(15,-10)(30,0)}
    \put(0,-30){\linethickness{0.2mm} \qbezier(0,0)(15,-10)(30,0)}
    \put(120,-30){\linethickness{0.2mm} \qbezier(0,0)(15,-10)(30,0)}
    \put(150,30){\line(1,0){60}} \put(180,0){\line(1,0){30}} \put(180,-30){\line(1,0){30}}
    \put(-90,-30){\line(1,0){60}} \put(-90,0){\line(1,0){30}} \put(-90,30){\line(1,0){30}}
  }
\end{picture}

The corresponding fat graph is

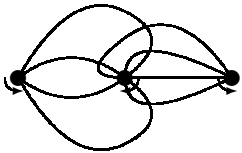
\begin{picture}(50,20)(-5,30)
  \put(0,0){
    \put(0,0){\circle*{6}}
    \put(0,0){\qbezier(-5,0)(-5,-5)(0,-5) \put(2,-5){\vector(1,0){0}}}
  }
  \put(40,0){
    \put(0,0){\circle*{6}}
    \put(0,0){\qbezier(5,0)(5,-5)(0,-5) \put(-2,-5){\vector(-1,0){0}}}
  }
  \put(80,0){
    \put(0,0){\circle*{6}}
    \put(0,0){\qbezier(-5,0)(-5,-5)(0,-5) \put(2,-5){\vector(1,0){0}}}
  }
  \linethickness{0.2mm}
  \qbezier(0,0)(20,15)(40,0)\qbezier(0,0)(20,-15)(40,0)
  \qbezier(0,0)(20,35)(40,25)\qbezier(0,0)(20,-35)(40,-25)
  \qbezier(40,25)(60,15)(40,0)\qbezier(40,-25)(60,-15)(40,0)
  \qbezier(40,0)(60,0)(80,0)\qbezier(40,0)(40,20)(80,0)\qbezier(40,0)(40,-20)(80,0)
  \qbezier(40,0)(20,0)(40,15)\qbezier(40,15)(60,30)(80,0)
\end{picture}
\be
\# \ \text{of vertices} = 3 = s_{||}\nn \\
\# \ \text{of edges} = 8 = n \nn \\
\# \ \text{of loops}= 1 = s_{=} \nn \\ \nn \\
\text{dim}
= \left([N][N-1]^2 + [N-2] [N] [N-1] \left ([2] + [2]^3 \right) \right) \
\stackrel{N=2,q=1}{\longrightarrow} \ 2 = 2^{\#\ \text{of loops}}
\ee

As we see, the graph has genus $3$, and coincides with Turaev genus of the 3-strand diagram.

Of course, it has nothing to do with Turaev genus of $8_{19}$ -- just provides an estimate
from above.
If we take the standard 4-strand representation of $8_{19}$ instead of the 3-strand one,
with $n=9$,  $s_{||}=4$, $s_{=} = 3$ we get Turaev genus $2$ -- and again this is exactly
the genus of the corresponding matrix-model fat graph.

The two last examples illustrate the common feature of all torus knots:
the two topologically equivalent closed-braid configurations $T_{[m,k]}$and $T_{[k,m]}$
where the first stands the number of links, have different Turaev genera and smaller is
the one where the number of strands is even (for knots $m$ and $k$ have different parities).
Indeed, for the $m$-strand torus closed braid $T_{[m,k]}$ we have:
\be
n = (m-1)k, \ \ \ \ \
s_{||} = m,  \ \ \ \ \ s_{=} =\left\{ \begin{array}{cccc} k &\text{if} &m & \text{is even} \\
1 &\text{if} & m &  \text{is odd} \end{array} \right.
\ee
Thus Turaev genus is either $\frac{(m-1)(k-1)}{2}$ if $m$ is odd, or
$\frac{(m-2)(k-1)}{2} =\frac{(m-1)(k-1)}{2} - \frac{k-1}{2} < \frac{(m-1)(k-1)}{2}$
if $m$ is even.
The same are genera of the corresponding {\it cross} fat graphs.


\section{Conclusion}

In this paper we described the base for promoting the hypercube approach of \cite{DM3}
to a practically working machinery for evaluation of HOMFLY polynomials.
It is now applicable for calculations within at least 15 intersections, what is not
only sufficient to reproduce most results on the fundamental HOMFLY in \cite{katlas},
but also to calculate double and even triple cables.
This is important, because the approach of \cite{DM3} is essentially different from the
powerful methods of \cite{Kaul,RTmod,AnoMcabling}, based on the Reshetikhin-Turaev
representation-theory and Witten's conformal-block formalisms
-- intersection exists only in the Jones case of $N=2$
(where it is provided by Kauffman's ${\cal R}$-matrix \cite{KauffR}).
In difference with those formalisms, however, the hypercube approach has wider applications:
it was originally introduced \cite{Kh} to study $\beta$-deformation \cite{betadefo} and
superpolynomials \cite{superpols} and was recently applied \cite{DM3virt} to virtual knots \cite{virt}.
Its version in \cite{DM3} is currently the only way to construct HOMFLY (not just Jones)
polynomials in the latter case (in the former a far more sophisticated but better developed
Khovanov-Rozansky matrix-factorization technique is available \cite{KhR}, at least in
principle).
However, since \cite{DM3} is far from providing rigorous formulations,
nothing is yet guaranteed in this approach -- even Reidemeister invariance.
One of the outcomes of our present work is a wide check that such invariance is indeed present,
moreover, we now almost understand, why.
Also provided in the present text are absolutely new formulas --
for cabled HOMFLY of the simplest virtual knots,
what now allows to study differential expansions \cite{diffexpan} and possibilities to introduce
the analogues of colored knot polynomials \cite{DM3virt}.
Another result is availability of practically any quantum dimensions, associated with the
hypercube vertices -- this should help in equally explicit construction of commuting morphisms
and thus complete the construction of a {\it practically working} alternative to
almost un-handleable matrix-factorization construction of Khovanov-Rozansky polynomials.

All this is achieved by developing the projector formalism of \cite{AnoMKhR} into a set
of recursions, very similar in spirit to matrix-model Ward identities (Virasoro constraints).
A matrix model itself is attached to the story at $q=1$, but it turns out that what survives
at $q\neq 1$ are recursions, associated exactly with Reidemeister moves(!).
The main purpose of the present text is to explain the very set of emerging ideas --
to taste the spirit of emerging theory, which is in many respects a little unusual, --
while various more technical considerations and applications will be provided elsewhere
(see \cite{MMP2} for the next step).

\section*{Acknowledgements}

A.M.'s gratefully acknowledge the hospitality and support of
the Simons Center for Geometry and Physics, Stony Brook University
at which some of the research for this paper was performed.
A.P. wants to thank Leiden Lorenz Center and Max Planck Institute
for Mathematics in Bonn for hospitality and wonderful scientific atmosphere.

Our work is partly supported by grants NSh-1500.2014.2, by RFBR grants   13-02-00478
(A.M.), 14-02-00627 (An.M.), 13-02-00457 (A.P.),
by the joint grants 15-51-52031-NSC-a, 14-01-92691-Ind-a,
and by 15-52-50041-YaF and also by the young-scientist's grant
14-01-31492 mol-a (A.P.). Also we are partly supported
by the Laboratory of Quantum Topology of Chelyabinsk State University
through Russian Federation government grant 14.Z50.31.0020 (An.M.),
by the Vici grant of the Netherlands Organization for Scientific Research (NWO) (A.P.)
and by the Brazilian Ministry of Science, Technology and Innovation through the National Counsel
of Scientific and Technological Development (A.M.).

\end{document}